\documentclass[aps,prb,twocolumn,superscriptaddress,longbibliography,noeprint]{revtex4-2}

\usepackage{graphicx}
\usepackage{bm}
\usepackage{multirow}
\usepackage[table]{xcolor}
\usepackage[export]{adjustbox}
\makeatletter
\let\old@makecaption=\@makecaption
\usepackage{subcaption}
\let\@makecaption=\old@makecaption
\makeatother
\usepackage{color}
\usepackage{orcidlink}
\usepackage{wasysym}
\newcommand\krc{K$_2$ReCl$_6$}
\newcommand\ksc{K$_2$SnCl$_6$}

\begin{document}

	\title{Structural studies on $A_2$ReCl$_6$ ($A$=K, Rb, Cs): absence of Jahn-Teller distortion}

\author{A. Bertin\,\orcidlink{0000-0001-5789-3178}}
\email{bertin@ph2.uni-koeln.de}
\affiliation{II. Physikalisches Institut, Universit\"at zu K\"oln, Z\"ulpicher Stra\ss e 77, D-50937 K\"oln, Germany}	

\author{L. Kiefer\,\orcidlink{0009-0008-5716-2816}}
\affiliation{II. Physikalisches Institut, Universit\"at zu K\"oln, Z\"ulpicher Stra\ss e 77, D-50937 K\"oln, Germany}	

\author{V. Pomjakushin\,\orcidlink{0000-0003-2180-8730}}
\affiliation{Laboratory for Neutron Scattering and Imaging, PSI, CH-5232 Villigen PSI, Switzerland}

\author{O. Fabelo\,\orcidlink{0000-0001-6452-8830}}
\affiliation{Institut Laue-Langevin, 71 avenue des Martyrs 38042 Grenoble, France}

\author{P. Becker\,\orcidlink{0000-0003-4784-3729}}
\author{L.~Bohat\'{y}}
\affiliation{Institute of Geology and Mineralogy, Sect. Crystallography, University of Cologne, 50674 Cologne, Germany}

\author{M.~Braden\,\orcidlink{0000-0002-9284-6585}}
\email{braden@ph2.uni-koeln.de}
\affiliation{II. Physikalisches Institut, Universit\"at zu K\"oln, Z\"ulpicher Stra\ss e 77, D-50937 K\"oln, Germany}

\begin{abstract}

K$_2$ReCl$_6$ belongs to the antifluorite family and exhibits a sequence of structural transitions above the onset of magnetic order at $T_{\rm N}=12$\,K. Because of its $5d^3$ electronic configuration in an octahedral coordination, the ground state is a pure spin state without orbital degeneracy within the $LS$ coupling scheme, but it can become Jahn-Teller active in the strong spin-orbit coupling limit described by the $jj$ coupling [S. Streltsov and D.\,I. Khomskii, Phys. Rev. X {\bf 10}, 031043 (2020)]. While the structural transitions in K$_2$ReCl$_6$ are understood in terms of octahedral rotation and tilting, the possible impact of a Jahn-Teller distortion remains an open issue. We report on comprehensive crystal-structure studies by means of powder neutron and single-crystal x-ray diffraction on K$_2$ReCl$_6$ and on K$_2$SnCl$_6$. The latter material is used as a reference, because it exhibits the same sequence of structural transitions as K$_2$ReCl$_6$, but possesses a filled $4d$ shell ruling out a Jahn-Teller distortion. While the ReCl$_6$ octahedron in K$_2$ReCl$_6$ presents sizable distortions at intermediate temperatures, there is no such distortion persisting to low temperatures excluding a sizable Jahn-Teller effect. Studies on polycrystalline samples of Rb$_2$ReCl$_6$ and Cs$_2$ReCl$_6$, in which the structural transitions are suppressed due to the larger alkaline ionic radius, also do not find any indications for a  Jahn-Teller distortion. 
\end{abstract}

\date{\today}

\maketitle

\section{Introduction}

Many recent developments in condensed matter physics result from the interplay between strong spin-orbit coupling (SOC) and electronic correlations, and much efforts have been undertaken on heavy $4d$ and $5d$ transition-metal compounds \cite{Raghu2008,Zhang2012,Maciejko2015,Trebst2022}.
For instance, Ir$^{4+}$ can develop a spin-orbit entangled effective moment $j_{\rm eff}=\frac{1}{2}$
with peculiar properties \cite{takagi2019}. In the honeycomb iridates, the magnetic interaction becomes anisotropic at least approaching the fully bond-directional magnetic interaction in the Kitaev model, whose
exact solution reveals a quantum-spin-liquid ground state~\cite{Kitaev06}.
The impact of strong SOC on other $5d$ electronic configuration has been much less studied.

 With this purpose, other materials for novel spin-orbit driven phenomena are synthesized. Among the $5d$ Mott insulators K$_2$ReCl$_6$ is a promising material due to its $5d^3$ electronic configuration in a cubic crystal field. While for lighter elements, the weak SOC $\lambda$ with regards to Hund's coupling $J_{\rm H}$, is well described within the Russell Saunders coupling scheme, the stronger SOC in heavier elements is best accounted for within the $jj$ coupling scheme \cite{Khomskii14}. In the former case, the resulting ground state in \krc ~would be a pure $S=\frac{3}{2}$ spin state without orbital degeneracy, while in the latter case, the resulting effective spin-orbit entangled moment would be $j_{\rm eff}=\frac{3}{2}$ and the ground state Jahn-Teller active~\cite{Streltsov20}. For transition metals in an octahedral environment, the $t_{2g}$ orbital degeneracy can be lifted through the coupling with vibration modes, namely an orthorhombic (called $Q2$ mode) and a tetragonal ($Q3$ mode) distortion~\cite{Goodenough63,Khomskii14}.
 In the intermediate SOC range, a compressed ReCl$_6$ octahedral geometry would be stabilized, but for larger $\lambda$, the Jahn-Teller distortion amplitude is reduced \cite{Streltsov20}.
 This situation is reminiscent of the $A_2$TaCl$_6$ antifluorite compounds with a $5d^1$ electronic configuration ~\cite{Ishikawa19}, where a cubic to tetragonal transition occurs. The elongated octahedral geometry is stabilized in the case of K$_2$TaCl$_6$, while a compressed one is stabilized for Rb$_2$TaCl$_6$ and Cs$_2$TaCl$_6$. The latter case is expected in the case of a $5d^1$ electronic configuration and a linear Jahn-Teller distortion~\cite{Streltsov20}. However, authors of Ref.~\cite{Khomskii21} pointed out that anharmonic effects, expected to be stronger for the smaller K ion than for Cs and Rb ions, could restore the elongated octahedral geometry, similarly to the $e_g$ case in half-doped manganites~\cite{Khomskii00}. Furthermore, density functional theory (DFT) calculations emphasize that the tetragonal $Q3$ mode, in particular its sign, is the key to stabilize either a pure $S=\frac{1}{2}$ or a spin-orbit entangled $j_{\rm eff}=\frac{3}{2}$ ground state in the $4d^1$ Nb antifluorite compounds; and tuning the tetragonal distortion via epitaxial strain would induce a quantum phase transition between these two ground states~\cite{Weng21}.

 In \krc , the Hund coupling $J_H$ is expected to be comparable to the strength of the SOC, and therefore the $jj$ coupling scheme can possibly apply. For instance, in the case of the double perovskite Ba$_2$YReO$_6$ (Re$^{5+}$), RIXS experiments reveal $J_H<\lambda$~\cite{Yuan17}, and in Ba$_2$MgReO$_6$, while the electronic correlations drive the charge quadrupolar order, the associated Jahn-Teller distortion plays a crucial role in the ground state stabilization~\cite{Soh24}. On the other side, recent RIXS experiments on K$_2$ReCl$_6$ yield a ratio $\lambda/J_{\rm H} \approx 0.6$, which in the theory of reference \cite{Streltsov2020} is not large enough to induce a sizable Jahn-Teller distortion~\cite{Warzanowski24}, but the interplay between the Jahn-Tellereffect and the structural phase transitions was left open.
 Also recent DFT calculations reveal a small Jahn-Teller distortion only when enhancing the SOC beyond the self-consistent
 value \cite{Du25}.

\begin{figure*}[t!]
	\centering
	\includegraphics[width=1.8\columnwidth]{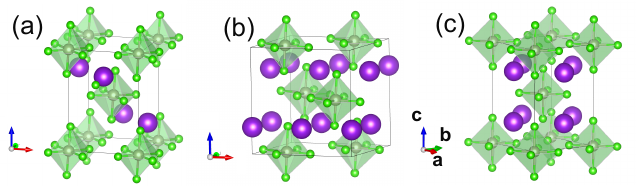}
	\caption{\label{struc} Crystal structure of \krc \ in the three low temperature
	phases as determined with PND. Panel (a) shows the structure at 15\,K in the monoclinic
	$P2_1/n$ phase. Panel (b) presents the monoclinic $C2/c$ phase at 90\,K. In these two monoclinic phases staggered rotation occurs around the $c_m$ axis and the in-phase tilt
	around the monoclinic $b_m$ axis that, however, changes between the two phases. Panel (c) shows the crystal structure in the tetragonal phase with only the staggered rotation
	around $c_{tet}$. Here we show the smaller lattice corresponding to $P4/mnc$. In all
	panels the red-green-blue arrows indicate the orientation of the lattice, K sites are
	shown in purple spheres and ReCl$_6$ octahedrons in green. Red, green and blue arrows indicate the $\boldsymbol{a}$, $\boldsymbol{b}$ and $\boldsymbol{c}$ lattice vectors.
	Pictures were drawn with the visualization software~\textsc{VESTA3}~\cite{Momma2011}. }
\end{figure*}

K$_2$ReCl$_6$ belongs to the antifluorite family and undergoes a series of structural transitions at $T_{\rm t}=111$\,K, $T_{\rm m1}=104$\,K, and $T_{\rm m2}=76$\,K towards a tetragonal phase with space group $P4/mnc$, a first monoclinic phase with space group $C2/c$, and a second monoclinic phase with space group $P2_1/n$, respectively \cite{Busey1962,Busey1962a,OLeary1970, Armstrong1980,Willemsen1977,Willemsen1977a,Smith1966,Minkiewicz1968,Bertin24a}.
Throughout this work we use the space-group symbols to label these phases.
These structural phase transitions are understood in terms of rotation and tilt of the ReCl$_6$ octahedrons (see illustrations in Fig. 1) and arise from bond-length mismatch similar to the analogous transitions in perovskites \cite{Bertin24b}. 
At $T_{\rm N}$=12\,K,  AFM long-range magnetic order sets in, associated with a weak structural distortion~\cite{Bertin24a,Smith1966,Minkiewicz1968}. 
However, the structural phase transitions severely impede the crytallographic analysis and so far no precise determination of the structural
parameters was reported for this material.
Therefore, it remains unclear whether there is a Jahn-Teller distortion driven through SOC in \krc ~ and how these structural distortions interfere with the rotation and tilting of the octahedron. The large rotation and tilting angles in the distorted low-temperature phases of \krc ~ make it more difficult to answer this question, because such distortions can induce a purely structural deformation of the transition-metal ligand octahedron, as it is observed in various perovskites \cite{Cwik2003,Khomskii14}. In order to distinguish electronically
and structurally driven octahedron deformations studying an isostructural compound with a completely filled $d$ shell appears most promising. K$_2$SnCl$_6$ exhibits the same sequence of structural phase transitions as K$_2$ReCl$_6$ \cite{Boysen76,Boysen78}, and as it has a fully occupied $4d^{10}$ electronic configuration, there is no Jahn-Teller effect rendering it an ideal reference compound.

With the aim to identify or exclude a Jahn-Teller driven structural distortion in $A_2$ReCl$_6$ ($A$=K, Rb, Cs),
we report comprehensive single-crystal x-ray diffraction (SXD) and  high-resolution powder neutron  diffraction (PND) measurements on K$_2$ReCl$_6$ and on K$_2$SnCl$_6$  in Sec.~III, where stoichiometry, local disorder, and the characterization of the low-temperature structural phases will be discussed. We also analyze the triclinic distortion accompanying the onset of magnetic order in K$_2$ReCl$_6$. In section IV we discuss the possible Jahn-Teller effect by comparing the crystal structure results for  K$_2$ReCl$_6$ and K$_2$SnCl$_6$. The absence of octahedral distortions at low temperature and the similar effects in both compounds
in the intermediate structural phases let us conclude that the Jahn-Teller effect is essentially inactive in \krc , although non-relaxed rotations and tilts can deform the structure.
In Sec.~IV we report on studies on polycrystalline samples of  Rb$_2$ReCl$_6$ and Cs$_2$ReCl$_6$, in which monovalent K$^+$ ions are replaced by larger Rb$^+$ and Cs$^+$ ions. Also in these materials we find no indication for a Jahn-Teller distortion.

\section{Experimental}
\label{methods_gen}

\subsection{Synthesis and characterization}

Single crystals of K$_2$ReCl$_6$ and K$_2$SnCl$_6$ were grown from HCl solution by controlled slow evaporation of the solvent \cite{Bertin24a,Bertin24b}. A commercial K$_2$ReCl$_6$ powder was purchased from Alfa Aesar, and we  crushed single crystals to obtain a powder K$_2$SnCl$_6$  sample. Detailed macroscopic measurements characterizing
the structural and magnetic phase transitions in our \krc ~ crystals are described in Ref. \cite{Bertin24a}. Measurements of the magnetization at 0.1\,T of the powder  \krc ~ sample using a quantum design MPMS-XL7 superconducting quantum interference device magnetometer
indicated a N\'eel temperature of 11.8(4)\,K in perfect agreement with the single-crystal studies \cite{Bertin24a}.

Powder samples of Rb$_2$ReCl$_6$ and Cs$_2$ReCl$_6$ were obtained by reaction of a solution of commercial K$_2$ReCl$_6$ with solution of RbCl, respectively CsCl, in diluted HCl. The precipitated microcrystalline Rb$_2$ReCl$_6$, respectively Cs$_2$ReCl$_6$, was washed several times with diluted HCl and dried.
Powder x-ray diffraction experiments as a function of temperature were performed in Bragg-Brentano geometry on a $Stoe$ diffractometer using Cu K$_{\alpha,1}$ radiation revealing pure phases. Samples were mounted in 
a He evaporation cryostat and a small amount of Si powder was mixed with the sample to determine 
temperature dependent instrumental parameters in particular the small sample displacement perpendicular
to the sample surface.
Magnetic properties of both compounds were measured using a quantum design MPMS-XL7 superconducting quantum interference device (SQUID) magnetometer. 

\subsection{SXD experiments}

SXD measurements were performed in a temperature range $30$\,K$\leq T \leq 300$\,K on a Bruker single-crystal diffractometer D8-Venture at the University of Cologne with Mo $K_\alpha$ radiation, $\lambda=0.71$\AA, and on a D8-Venture at the Institut Laue-Langevin (ILL) with Ag $K_\alpha$ radiation, $\lambda=0.56$\AA. Note that the use of a shorter wavelength reduces the x-ray absorption. Samples were mounted using Cargylle oil on a Mitegen cryoloop. Integration was performed with the \textit{Apex4} software. Absorption correction and scaling were performed with the {\it Multiscale} algorithm. The sample sizes were described by a parallelepiped with dimensions reported in the supplemental material \cite{supplmat}. The D8-Venture diffractometers use a microfocus x-ray source equipped with multilayer optics, which monochromatize the beam and suppress higher-order harmonics. 
For the low temperature experiments we employed a nitrogen and a helium cryostream cooler for temperatures above and below 80\,K, respectively.  
The quality of the collected data sets is evaluated by the internal reliability values $wR^2$(int). Refinements were carried out with \textit{Jana2020}~\cite{Jana} applying an extinction correction with an isotropic Becker-Coppens formalism~\cite{Becker74}. 
%We use the ionic form factors of K$^{+}$, Sn$^{4+}$ and Cl$^{-}$, and the atomic form factor for Re because the ionic form factor of Re$^{4+}$ is not available. 

While data sets collected in the cubic high-temperature structure described with space group $Fm\bar{3}m$ can be integrated with the cubic face-centered unit cell, data sets collected in all lower-symmetry phases were integrated with a pseudo-cubic primitive unit cell, with lattice parameter $a\sim a_{cubic}$; see Tab.~S1 in the supplemental material \cite{supplmat}. Inherent to the symmetry reduction at the structural phase transitions, ferroelastic twin domains appear with up to 12 different orientations \cite{Bertin24a} that mimic the higher symmetry. Since the lattice
distortions remain small~\footnote{$\beta_{\rm m1} \approx \beta_{\rm m2} \approx 90^{\circ}$, $a_{\rm m2} \times \sqrt{2} \approx b_{\rm m2} \times \sqrt{2} \approx c_{\rm m2}$, $a_{\rm m1} \approx b_{\rm m1} \approx c_{\rm m1}$, and $ a_{\rm t} \times \sqrt{2} \approx c_{\rm t}$, where the indices $m2$, $m1$, and $t$ refer to the lattice parameters of the $P2_1/n$, $C2/c$, and $P4/mnc$ structural phases} one cannot separate the contributions of individual orientations, which sometimes is even intrinsically impossible. 
Therefore, the integrated datasets correspond to an incoherent superposition
of the up to 12 twin volumes, which was implemented in the refinements.

In the high-temperature cubic phase $Fm\bar{3}m$, the atomic displacement parameters (ADP)~\cite{Trueblood96} of $Me$ = Re, Sn and K ions stay isotropic, while two anisotropic ADPs are introduced for the Cl ions: $U_{\parallel}$ parallel and $U_{\perp}$  perpendicular to the $Me$-Cl bond. For the refinements in the lower-symmetry phases, the same two ADPs were refined for all Cl sites and ADPs of $Me$ and K ions were kept isotropic. 
% $R$-values and GOF parameters attesting the quality of the refinements are tabulated in Tab.~\ref{Data_set_all_KRC}. 
In order to calculate the bond distances and rotation and tilt angles of the octahedron, we combined the atomic positions obtained from SXD with the lattice parameters obtained from powder x-ray diffraction and documented in Ref.~\cite{Bertin24a} for K$_2$ReCl$_6$. 
For K$_2$SnCl$_6$ the lattice parameters are obtained from our high-resolution PND experiments. The crystal-structure refinements yield very good reliability factors and the results are resumed in Tables S2 and S3 in the supplemental material~\cite{supplmat}.

\subsection{PND experiments}

High-resolution PND experiments were performed on the HRPT diffractometer at the SINQ neutron source of the Paul Scherrer Institute. 5.5\,g of a K$_2$ReCl$_6$ sample  and 5.3\,g of K$_2$SnCl$_6$ sample were filled in 8\,mm diameter $\times$ 50\,mm height cylindrical Vanadium sample cans that were mounted in an ILL-orange-type cryostat. A wavelength of 1.49\,\AA~was used for both samples. K$_2$SnCl$_6$ patterns were collected in {\sl high-intensity} mode, i.e.\ with a $40^{\prime}$ primary collimator, and K$_2$ReCl$_6$ patterns were collected in {\sl medium resolution} mode, i.e.\ with a $12^{\prime}$ primary collimator. 
In the experiment on the \ksc ~ sample for temperatures around 260\,K, the determined temperature on the sample stick slightly deviated from the 
sample temperature, $\Delta$T$\sim$3\,K, due to insufficient thermalization between the sample and the sonde and is labelled $T^\prime$ throughout the paper. 
Rietveld refinements were performed with the \textsc{Fullprof} program package~\cite{Carvajal93}. An instrument resolution file reflecting the exact geometry of the experimental setup, modeled with a Thomson-Cox-Hastings (TCH) pseudo-Voigt function~\cite{Thompson87} convoluted with an axial-divergence asymmetry function~\cite{Finger94}, was used to describe the peak profile for \krc , and a microstructure analysis has been applied.
Background was modeled with a cosine Fourier series of $6^{\rm th}$ degree and an absorption correction was introduced \footnote{The absorption correction is $\exp(-\Sigma R)$, where $R$ is the radius of the cylindrical sample and $\Sigma$ is the total absorption length given by
$ \Sigma=\frac{N_f}{v_0}f\frac{\lambda}{1.8} \sum_i c_i \sigma_{a,i}$,
where $N_f$ is the number of formula unit, $v_0$ the volume of the unit cell at 120\,K, $f=0.7$ the estimated powder filling, $\lambda=1.494$\AA~ the neutron wavelength (normalized to 1.8\AA), $\sigma_{a,i}$ the absorption cross section~\cite{cross_sections} of element $i$ contained $c_i$ times per formula unit.}\cite{cross_sections}.
For \krc ~ the powder-diffraction pattern measured in the cubic phase at 120\,K was used as a reference pattern in order to determine additional instrumental parameters (zero position of the detector, the sample displacement and transparency, and one additional parameter accounting for the peak asymmetry). Thereafter, these parameters were fixed in the refinements with data obtained at lower temperature.

The same ADPs methodology~\cite{Trueblood96} as used for the SXD analysis was applied~\footnote{Note that in FullProf, the anisotropic ADPs are computed with the dimensionless $\beta_{ij}$ parameters where the indices $i,j$ refer to the crystallographic axis. 
In the manuscript, $U_{ij}$ (in~\AA$^2$) are reported and the following transformation is used: $U_{ij}=\beta_{ij}/(2 \pi^2 a^{\star}_ia^{\star}_j)$, where $a^{\star}_i$ denote the reciprocal lattice parameters~\cite{Trueblood96}. Note that when setting the Cl ADPs constraints to refine only 2 parameters, only the change of basis is considered in the $\beta$ calculations, but not the small tetragonal/monoclinic splitting and monoclinic angles, the latter having only a minor impact of the order $10^{-4}$ on the $U_{ij}$ values, well below the standard deviations.}. 
Table I resumes the results of the PND refinements and presents the positions of the sites in the different phases.

The average apparent particle size was refined with the reference pattern at 120\,K, and an almost resolution limited grain size with diameter $D=1611.1(9)$\AA~is found for \krc ~ and fixed for all other temperatures (for \ksc ~  $D=1728(2)$\AA~is obtained with ambient-temperature data).
The strain was refined for all patterns and is plotted in Fig.~S1 in the supplemental material~\cite{supplmat}. Our results indicate an almost temperature independent behavior, and the still small value inferred from the Rietveld refinement in the \krc ~ cubic phase at 120\,K $e=4.768(5)\times 10^{-4}$\permil~ is about five times larger than the value deduced at room temperature in K$_2$SnCl$_6$~\cite{supplmat}.

\begin{table*}[t!]
		\centering
		\caption{\label{KRC_HRPT_refinement_1} Structural parameters inferred from Rietveld refinements with PND data on K$_2$ReCl$_6$  (upper part) and on \ksc ~(lower part) measured in the cubic $Fm\bar{3}m$, in the tetragonal $P4mnc$, in the first monoclinic $C2/c$ and in the second monoclinic $P2_1/n$ phases. Atomic positions in the $C2/c$ phase are corrected for the origin shift ($\frac{1}{4},~\frac{1}{4},~0$). \krc ~ data at 14\,K and 12\,K are very similar to the ones at 15\,K and not reported in this table for clarity. $R_p$, $R_{wp}$, and $R_{exp}$ are the profile, weighted-profile, and expected profile $R$ factors corrected for background and expressed in~\%. $\chi^2$ denots the goodness of fit (GOF).
			 $T^{\prime}$ refers to the measured temperature in the experiments on \ksc . All ADPs $U$ are given in \AA$^2$, lattice angles in degrees and lattice constants in \AA . The structural parameters determined by single-crystal diffraction can be found in the supplemental material \cite{supplmat}.}
\vspace{4pt}
\scalebox{0.90}{		
		\begin{tabular}{l|lllll}
			\hline
	T		&120\,K&107\,K&90\,K&15\,K&1.5\,K\cr
	\krc	& $Fm\bar{3}m$&$P4/mnc$&$C2/c$&$P2_1/n$&$P_S\bar{1}$ \cr
			\hline
	$a$ $\alpha$ & $a_{\rm c}$=9.7629(1)&$a_{\rm t}$=6.89512(3) &$a_{\rm m,1}$=9.72660(7)            &$a_{\rm m,2}=$6.87995(4)          &
			$a_{\rm t}$=6.8789(3)    $\alpha_{\rm t}$=89.986(3)  \cr
		$b$ $\beta$		&					  &                       &$b_{\rm m,1}$=9.76112(9) $\beta_{\rm m,1}$=90.043(2)            &$b_{\rm m,2}$=6.85975(3)  $\beta_{\rm m,2}$=90.102(7)          &$b_{\rm t}$=6.8564(3)    $\beta_{\rm t}$=90.110(1)  \cr	  
		$c$ $\gamma$		&					  &$c_{\rm t}$=9.77059(9) &$c_{\rm m,1}$=9.77322(9)            &$c_{\rm m,2}$=9.77919(5)           &$c_{\rm t}$=9.78321(4) $\gamma_{\rm t}$=89.916(3) \cr
			\hline
			U$_{\rm iso}$(Re)&0.0056(2)&0.004(3)&0.0055(3)&0.0021(2)&0.0017(2) \cr			
			
			K & ($\frac{1}{4}$,$\frac{1}{4}$,$\frac{1}{4}$) & ($\frac{1}{2}$,0,$\frac{1}{4}$) & ($\frac{1}{4}$,0.266(1),$\frac{1}{4}$)/($\frac{1}{4}$,-0.245(2),$\frac{1}{4}$)  &{0.5044(8),0.9843(7),0.2509(10)} & {0.5051(7),0.9842(6),0.2492(7)}\cr
			U$_{\rm iso}$(K)&0.0200(4)&0.0168(6)&0.0118(10)&0.0099(5)&0.0095(4) \cr	
			
			Cl$_1$ &0.24024(5),0,0&0.2426(4),-0.2344(4),0&0.2415(3),-0.0133(3),-0.0134(4)&0.2606(2),-0.2214(2),-0.0157(2)& 0.2605(2),-0.2208(2),-0.0158(2)\cr
			Cl$_2$   &0, $x$, 0&-$y$,$x$,0&0.0108(4),0.2398(3),-0.0032(8)&0.2207(2),0.2616(2),-0.0108(2)&0.2212(2),0.2614(2),-0.0110(2)\cr
			Cl$_3$   &0, 0, $x$&0, 0, 0.2428(3)&0.0153(3),-0.0002(5),0.2406(4)&0.0271(2),-0.0011(3),0.2400(2)&0.0271(2),-0.0012(3),0.2397(1)\cr				
			U$_{\parallel}$(Cl)&0.0094(3)&0.0069(3)&0.0071(5)&0.0036(1)&0.0039(3) \cr	
			U$_{\perp}$(Cl)&0.0272(2)&0.0295(3)&0.0192(3)&0.0091(2)&0.0081(2) \cr					
			\hline 	
			$R_p$, $R_{wp}$ &9.65~8.01&12.9~11.2&10.1~9.65&6.78~7.23&4.87~5.39 \cr
			$R_{exp}$, $\chi^2$ &5.52~2.1&5.12~4.8&5.32~3.3&4.79~2.3&3.02~3.19 \cr	
			
			\hline		
		\end{tabular}		
	}
	\vspace{8pt}
	
\scalebox{0.90}{				
		\begin{tabular}{l|llll}
			\hline
			$T^{\prime}$&255\,K&160\,K&40\,K&1.5\,K\cr
	\ksc	&$C2/c$&$P2_1/n$&$P2_1/n$&$P2_1/n$	\cr
			\hline
			
	$a$   &$a_{\rm m,1}$=9.9586(3)&$a_{\rm m,2}$=7.0181(1)&$a_{\rm m,2}$=6.98310(6)&$a_{\rm m,2}$=6.98098(6) \cr
	$b$		&$b_{\rm m,1}$=9.9722(3)&$b_{\rm m,2}$=7.0117(1)&$b_{\rm m,2}$=7.01198(5)&$b_{\rm m,2}$=7.01398(5) \cr	
	$c$  	&$c_{\rm m,1}$=10.0123(2)&$c_{\rm m,2}$=9.9848(5)&$c_{\rm m,2}$=9.92215(9)&$c_{\rm m,2}$=9.91774(9) \cr	
			$\beta$  	&$\beta_{\rm m,1}$=90.053(6) &$\beta_{\rm m,2}$=90.202(2) &$\beta_{\rm m,2}$ 90.341(1) &$\beta_{\rm m,2}$=90.3534(7) \cr	
			\hline
			$U_{\rm iso}$(Sn)&0.0191(9)&0.0121(8)&0.0048(5)&0.0025(5) \cr			
			
			K & ($\frac{1}{4}$,0.270(3),$\frac{1}{4}$)/($\frac{1}{4}$,-0.242(3),$\frac{1}{4}$) & {-0.011(1),0.4728(9),0.2502(15)} & {-0.0093(6),0.4625(5),0.2515(7)} & {-0.0098(7),0.4621(5),0.2520(7)}\cr
			$U_{\rm iso}$(K)&0.039(2)&0.021(1)&0.0083(7)&0.0065(7) \cr	
			
			Cl$_1$ &0.2424(7),-0.0219(13),-0.0151(8)&0.2708(5),-0.2128(6),-0.0214(4)&0.2749(2),-0.2076(3),-0.0271(2)&0.2756(2),-0.2074(3),-0.0277(2)\cr
			Cl$_2$   &0.0212(13),0.2405(7),-0.0037(16)&0.2096(5),0.2741(6),-0.0186(4)&0.2060(2),0.2772(2),-0.0222(2)&0.2057(2),0.2765(2),-0.0228(2)\cr				
			Cl$_3$   &0.0134(8),-0.0019(9),0.2406(4)&0.0423(4),0.0007(6),0.2409(2)&0.0505(2),0.0014(3),0.2418(1)&0.0505(2),0.0007(3),0.2422(2)\cr					
			$U_{\parallel}$(Cl)&0.0220(9)&0.0132(2)&0.0046(2)&0.0049(2) \cr	
			$U_{\perp}$(Cl)&0.0551(8)&0.0269(4)&0.0090(2)&0.0054(4)   \cr					
			\hline 	
			$R_p$, $R_{wp}$ &15.1~11.7& 9.84~9.67&6.45~7.15&6.81~7.56 \cr
			$R_{exp}$, $\chi^2$ &10.6~1.22&8.42~1.32&6.65~1.16&6.12~1.53 \cr	
		
		\end{tabular}	
	}
	\end{table*}

\section{Temperature dependence of the crystal structure in {K$_2$R\lowercase{e}C\lowercase{l}$_6$}  and in {K$_2$S\lowercase{n}C\lowercase{l}$_6$}}
\label{SC_XRD}

%We report the results of comprehensive SXD experiments performed on three diffractometers with  five K$_2$ReCl$_6$ single crystals over a broad temperature range covering the four structural transitions in this material.
%Reproducibility of the data from various samples, and measured on different instruments, is a key to strengthen our conclusions. For each measurement, an overview of the data set completeness is given in Tab.~\ref{Data_set_all_KRC}.   	
%

%At room temperature, both materials crystallize in the common antifluorite K$_2$PtCl$_6$ structure~\cite{Ewing28}, a $fcc$ cubic structure described with the $Fm\bar{3}m$ space group, where the $Me$, K and Cl ions occupy the Wyckoff positions $4a$ (0,0,0), $8c$ $(\frac{1}{4},\frac{1}{4},\frac{1}{4})$, and $24e$ ($x$,0,0), respectively.

\subsection{Stoichiometry of  {K$_2$ReCl$_6$} and  {K$_2$SnCl$_6$}}

The stoichiometry of the investigated samples was verified using SXD data sets collected in the high-temperature phase and by keeping the site occupancy of Re fixed to one.  The occupancies of the K and Cl ions are refined and are tabulated in  Tab.~S4 in the supplemental material~\cite{supplmat}.
%The $R$-values and GOF parameters for all reflections also tabulated on the left and right side of the slash symbol "/" correspond to harmonic refinements with stoichiometric and non-stoichiometric conditions, respectively.
With the exception of the K$_2$SnCl$_6$ crystal showing a slight excess of scattering at the Cl site, all samples do not exhibit any occupational deviation. Therefore, in the analyses described below, all occupation values were fixed to nominal stoichiometry.
\subsection{Study of the low-temperature phases in \krc }
\label{SC_XRD_results}

The SXD data completeness, and the $R$-values and GOF parameters documenting the quality of the refinements, are tabulated in Tab.~S1 in the supplemental material \cite{supplmat}, and the refined atomic positions, ADPs, and twin fractions are presented in Tab.~S2 in the supplemental material \cite{supplmat}.
Because of the large number of free parameters, in particular due to the presence of twin domains in the low-temperature phases, and despite constraining the ADPs to the symmetry of the cubic parent phase, some additional constraints were applied to avoid strong correlations ($>90$\%) between refined parameters. 
In all refinements in the tetragonal phase, we fix the ADP $U_{\parallel}$(Cl) to a value linearly scaled from the value found in the cubic phase at 120\,K for samples \#S2 and \#S3. 
This allows us to refine the volumes of the three domains, which are found to be nearly equipopulated for all samples. In the $C2/c$ phase, a constraint between the split K atomic positions is applied ($y({\rm K}_2)=y({\rm K}_1-0.5)$) for sample \#S2 in the refinements at 90\,K.
%\sout{and for sample \#S4 at 96\,K. However, for the latter, one {\bf which?} strong correlation persists.  Note that for this instrument, a vertical oscillatory motion of the sample reduced the precision of the observed structure factors. BETTER kick out S4 completely????}
Finally, for sample \#S2 at 80\,K, applying a resolution cut-off $\sin(\theta)/\lambda=0.9$\AA$^{-1}$~is enough to suppress any strong correlation.

\begin{figure}[h!]
	\centering
	\includegraphics[width=0.9997\columnwidth]{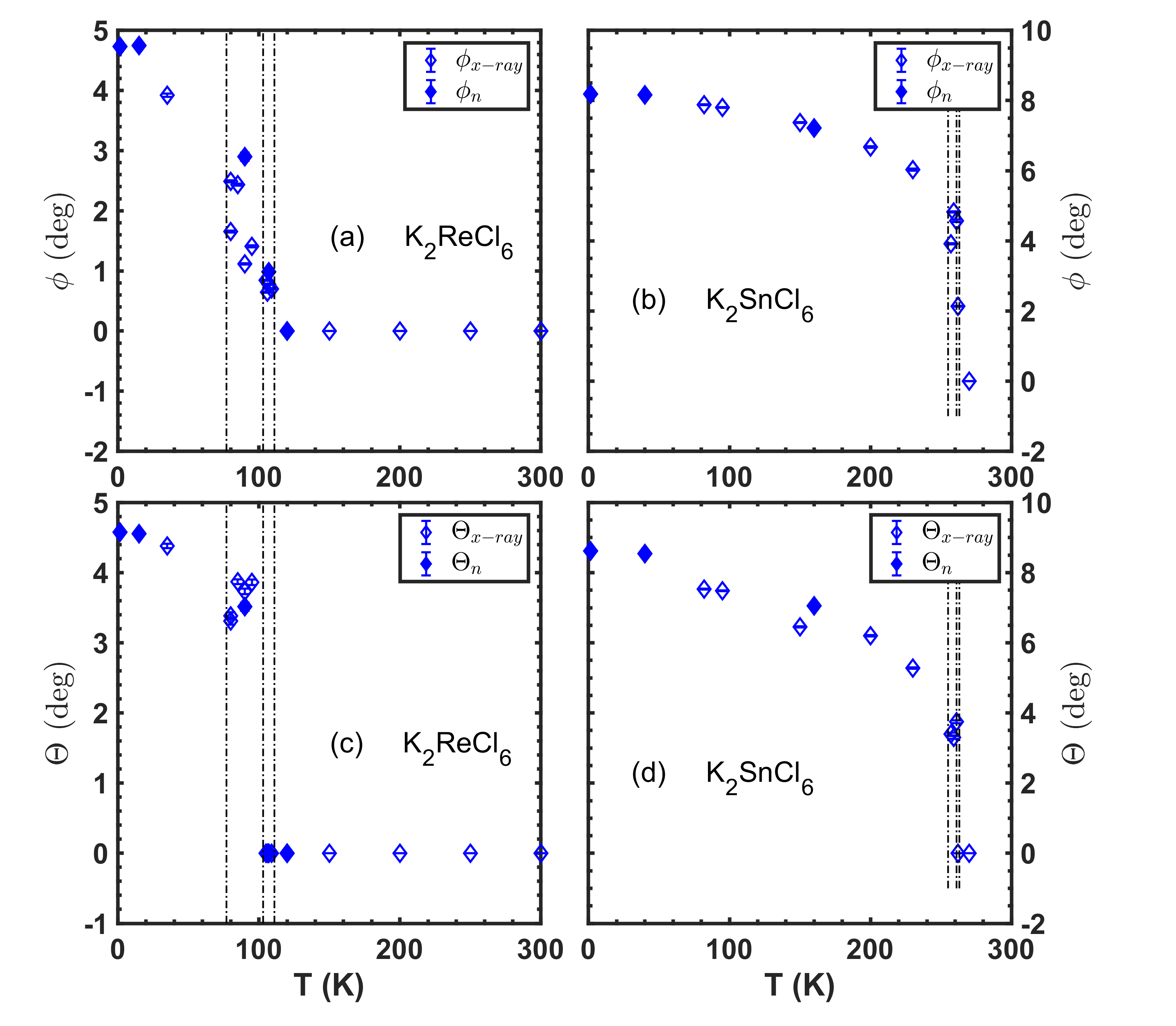}
\vskip-0.14cm
	\includegraphics[width=0.9997\columnwidth]{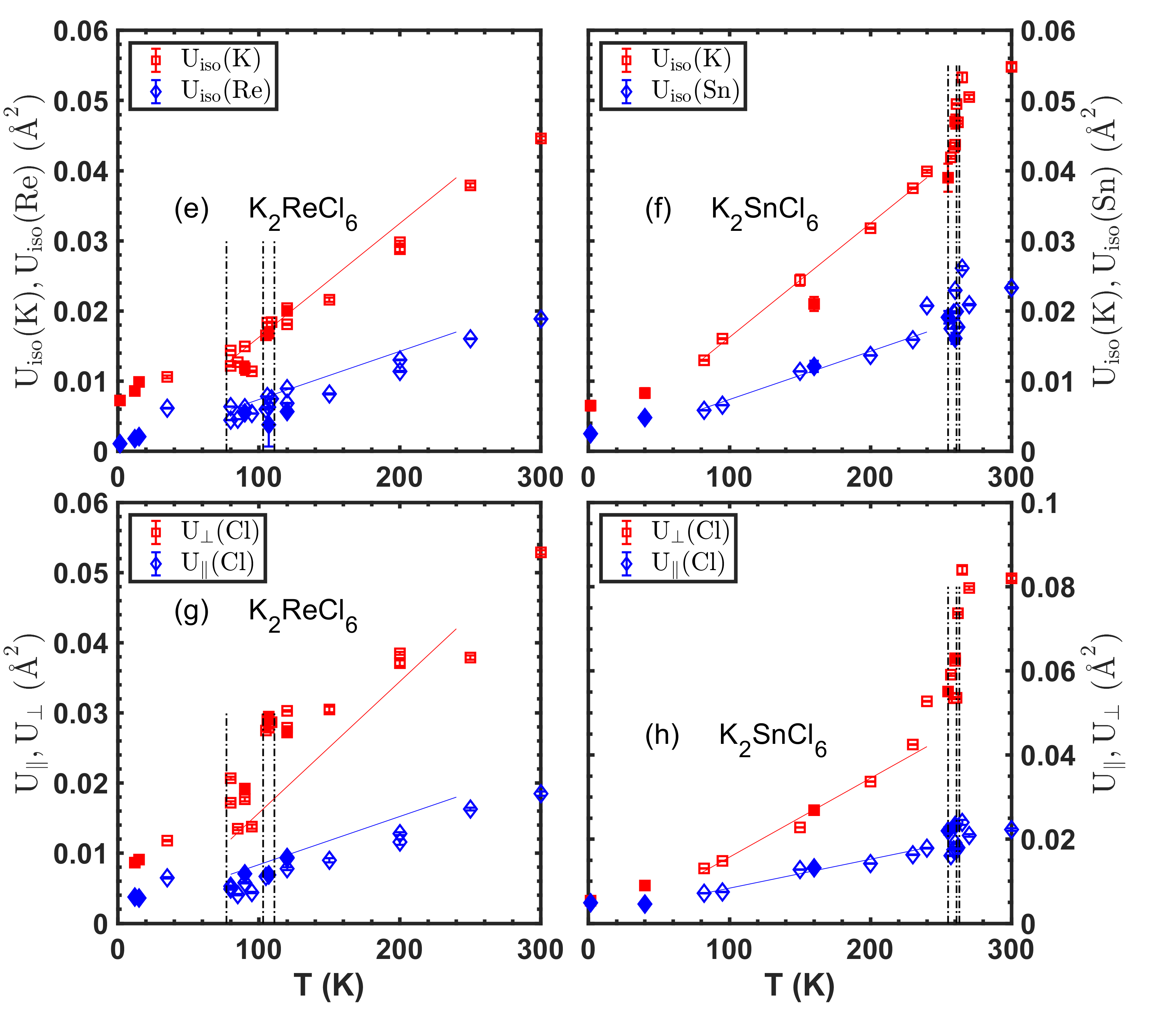}
	\caption{\label{KRC_KSC_all} The average octahedral rotation angle $\phi$, as defined in the main text, is given in panel (a) for K$_2$ReCl$_6$, and in (b) for K$_2$SnCl$_6$. Tilt angles $\theta$ are shown in panels (c) and (d). Note that in the $P2_1/n$ phase, the tilt axis is no longer parallel to a cubic [100] but parallel to cubic [110]. Panels (e) to (h) present the ADPs for the Re (left) and Sn (right) compounds. For K and Re/Sn the ADP is
	isotropic (e,f), while for Cl the two ADPs  $U_{\parallel}$ and $U_{\perp}$ are given in (g,h). The vertical dashed black lines indicate the structural transitions at $T_{\rm t}=111$~K and $T_{\rm m1}=103$~K, and $T_{\rm m2}=77$~K for
	\krc , and at 
	$T_{\rm t}=263$\,K, $T_{\rm m1}=260$\,K, and $T_{\rm m2}=$255\,K for \ksc .
	The solid lines are linear fits, used as a guide to the eye. Open and filled symbols refer to SXD or PND measurements, respectively.}
\end{figure}

K$_2$ReCl$_6$ undergoes a first structural transition towards a tetragonal phase with space group $P4/mnc$. 
Refinements are carried out with the non-standard notation $C4/mcg$ space group, which preserves the cubic unit cell of about 10\,\AA ~ length.
The cubic-to-tetragonal transition corresponds to a rotation of the ReCl$_6$ octahedron around
the tetragonal axis that is parallel to a Re-Cl bond. The rotation senses (clockwise or anticlockwise) are equal within
any plane perpendicular to the $c$ axis but the staggering is anti-phase parallel to it. Therefore, this distortion is associated with a mode at the Brillouin-zone boundary ($X$ mode of the $Fm\bar{3}m$ cell) and breaks
translation symmetry~\cite{Bertin24a}. The rotation angle, $\phi$, is calculated from the atomic positions of the Cl$_1$ and Cl$_2$  sites defining the octahedral basal plane (perpendicular to $c$). Note that Cl$_1$ and Cl$_2$ are symmetry related in $P4/mnc$ but not in $C2/c$ and $P2_1/n$. In the two monoclinic phases, the rotation angle $\phi$ is determined as the average of the rotations calculated using the two inequivalent basal Cl$_1$ and Cl$_2$ sites.
The transition from tetragonal to the first monoclinic phase $C2/c$ is associated with the rotary phonon mode softening at the Brillouin zone center ($\Gamma$ mode): an in-phase octahedral tilt against the monoclinic axis $b_{\rm m1}$, where $b_{\rm m1}$ is parallel to a cubic [100] direction.
Throughout this article we distinguish "rotation" and "tilt" for the $\boldsymbol{c}$ and $\boldsymbol{a,b}$ axes, respectively.
We determine the tilt angle $\theta$ by averaging the values calculated using the apical Cl$_3$ ion and the basal Cl$_1$ ion sitting on the $a_{\rm m1}$ axis.  In the second monoclinic phase $P2_1/n$, the tilt axis changes to $b_{\rm m2}$ parallel to a [110] cubic axis. The new tilt angle $\theta=\theta_{\rm basal}$ is estimated from the average tilt angles of the two basal Cls. The tilt $\theta=\theta_{\rm apical}$ is also calculated from the the apical Cl$_3$ ion, but differs only very little from $\theta_{\rm basal}$. 
These octahedral rotation and tilt angles are plotted against temperature in Fig.~\ref{KRC_KSC_all}.

The monoclinic low-temperature phases of \krc ~ thus combine a rotation and a tilt
distortion with distinct stacking schemes, which resembles the GdFeO$_3$ structure
type that is observed in numerous perovskite oxides and that
permits various distortions \cite{Khomskii14,Cwik2003,Koteras25}.
The complexity of the low-temperature $P2_1/n$ structure in the antifluorites \krc ~ and \ksc ~ is, however, considerably enhanced compared to the GdFeO$_3$ structure type, because
the in-plane Cl sites split. The antifluorites can be considered as double perovskites
$A_2BB'X_6$ with an empty $B'$ site, which represents a symmetry lowering compared
to an ideal perovskite. Therefore, the tilting emerging out of the tetragonally distorted structure
generates more degrees of freedom. Besides the main tilting that is in-phase at all sites
and that occurs around the monoclinic $b_{m1}$ and $b_{m2}$ axes in the two distinct
monoclinic phases, tilting is also possible around $a_{m1}$ and $a_{m2}$, respectively.
This tilting exhibits the same staggering as the rotation around $c$. 
These tiltings, however, are found to be smaller than 1$^{\circ}$ at all temperatures, supporting our assignment of the main octahedral rotation schemes across all structural transitions.

In the tetragonal phase, the rotation angle $\phi$ amounts to about 1$^{\circ}$, which disagrees with the order parameter deduced from nuclear quadrupole resonance (NQR) measurements, i.e.\  $\phi \approx 3^{\circ}$~\cite{Brown73}, but explains the weak superstructure reflections documented in Ref.~\cite{Bertin24a}.
The weak distortion also explains the necessity to constrain the refinement model in order to suppress the strong persistent correlation. The rotation angle $\phi$ increases continuously across the different structural phases up to $\phi=4.7(1)^{\circ}$ at the lowest temperature, see Fig.~2(a).
The discontinuity of the tilt angle $\theta$ at 104\,K would suggest a first-order character of the structural transition from the tetragonal $P4/mnc$ phase to the first monoclinic $C2/c$ phase but dense data close to this transition are lacking. 
We found a little temperature dependent angle of $\theta \approx 4^{\circ}$ in the $C2/c$ and $P2_1/n$ monoclinic phases.

The temperature dependencies of the two Cl ADPs  and of the isotropic ADPs of Re and K are plotted in Fig.~\ref{KRC_KSC_all} as a function of temperature. First, there is a good agreement between the ADPs determined with the SXD data on four different samples and the PND results.
% measured on the D8-Venture diffractometers~\footnote{The data from sample \#S4 measured on %the X8-Apex are not reported. Indeed, the ADPs values are much larger in all structural %phases; see Tab.~\ref{KRC_refinement_S4}, as a result likely of the instrument artifact %broadening the Bragg reflections}.
As commonly observed in the antifluorite materials~\cite{Bertin24b}, we found at high-temperatures $U_{\parallel} \ll U_{\perp}$ and quite large $U_{\perp}$  values. However, all ADPs shrink with temperature following mainly a linear behavior. 
The Cl ADPs become almost isotropic at the lowest temperature, indicating that the large high-temperature $U_{\perp}$ value has mainly a dynamical origin. Note however, that prior to the first cubic to tetragonal transition, $U_{\perp}$ deviates from the expected linear behavior, before significantly shrinking in the monoclinic phase $C2/c$.

PND patterns were collected in each of the crystallographic phases; see Tab.~\ref{KRC_HRPT_refinement_1} for the results of Rietveld refinements.
The lattice parameters are reported in Fig.~S2 in the supplemental material~\cite{supplmat}, and are compared with x-ray diffraction data reproduced from Ref.~\cite{Bertin24a}, showing good agreement, even though the $c$ lattice parameter may be slightly underestimated by the neutron data in the intermediate monoclinic phases. The PND data allow us to verify that
the signs of tetragonal or monoclinic splittings in the low-temperature phases were correctly attributed, see the discussion in the supplemental material~\cite{supplmat}. This it not trivial as the local distortion of the octahedron can overcompensate the effect of rigid octahedral rotation even in the case of corner-sharing perovskites~\cite{Koteras25}.

The ADPs and octahedral rotation/tilt angles deduced from PND Rietveld analysis are tabulated in  Tab.~\ref{KRC_HRPT_refinement_1} and plotted together with SXD results in Fig.~\ref{KRC_KSC_all}, showing a good agreement between the two techniques.

\subsection{Possible local disorder in \krc}
	%\label{disorder}
Raman scattering experiments show already at room temperature a violation of the cubic selection rules, and the persistence of a continuum across all structural transitions down to 5\,K. These observations were interpreted as the signature of octahedral rotational disorder breaking the local cubic symmetry~\cite{Stein23}. 
SXD experiments analyzing integrated Bragg reflection intensities determine the average long-range crystal structure, 
but nevertheless local disorder or local distortion possess a measurable impact.
The local disorder adds to the temperature-dependent phonon contribution to the ADPs and must lead to unusually large values \cite{Dunitz88,Braden2001,Bertin24b}. Pronounced disorder in the high symmetry phase must even change the probability density from a 3D Gaussian distribution to functions with several or extended maxima that can be described by higher polynomials~\cite{Kuhs88}, as it has
been demonstrated for the ferroelectric phase transition in K(H/D)$_2$PO$_4$~\cite{Nelmes82}.

In the $A_2MeX_6$ antifluorite compounds, strongly anisotropic ADPs of the $X$ ion are a common feature: $U_{\parallel} \ll U_{\perp}$  \cite{Bertin24b}. This purely dynamical feature arises from the peculiar arrangement allowing
for almost independent local rotations in $A_2MeX_6$ and is a hallmark of the rotational instability in the antifluorite family~\cite{Bertin24b}.
	
SXD measurements in the cubic $Fm\bar{3}m$ phase have been performed for K$_2$ReCl$_6$ (sample \#S1 at 250\,K, 200\,K, and 150\,K; samples \#S2 and \#3 at 120\,K) see Tab.~S1 in the supplemental material \cite{supplmat}, and the results of refinements carried out with both harmonic and anharmonic ADPs are compared in Tab.~S5 in the supplemental material~\cite{supplmat}. The anharmonic refinements improve only very little the $R$-values at all temperatures, and do not lead to a significant reduction of $U_{\perp}$(Cl), see Tab.~S5 in the supplemental material~\cite{supplmat}. 
Also refinements of a so-called split model~\cite{Nelmes82}, in which the Cl position is displaced perpendicular to the lattice axes, do not yield a significant improvement 
of the description~\cite{supplmat}, which is exceptionally good as indicated by the very low reliability factors. T

In conclusion, refinements with anharmonic ADPs in the cubic phase do not indicate the presence of strong rotational disorder in K$_2$ReCl$_6$ at any temperature in the cubic phase. Furthermore, at low-temperature the Cl ADP transversal to
the bond, U$_\perp$ [see Fig. 2(g)], is much smaller than the square of the static displacement of the Cl due to the
rotation. This documents well defined long-range rotational distortions with only a minor
role of local disorder that, however, for such ferroelastic phase transitions always exists due to the domain walls. As it is discussed in reference \cite{Bertin24b} for osmates and iridates, also the large ADPs in \krc \ stem mostly from the peculiar dynamics of the materials with soft phonons  spread over a large part of the Brillouin zone.

\begin{figure}[h!]
	\centering
	\includegraphics[width=0.95\columnwidth]{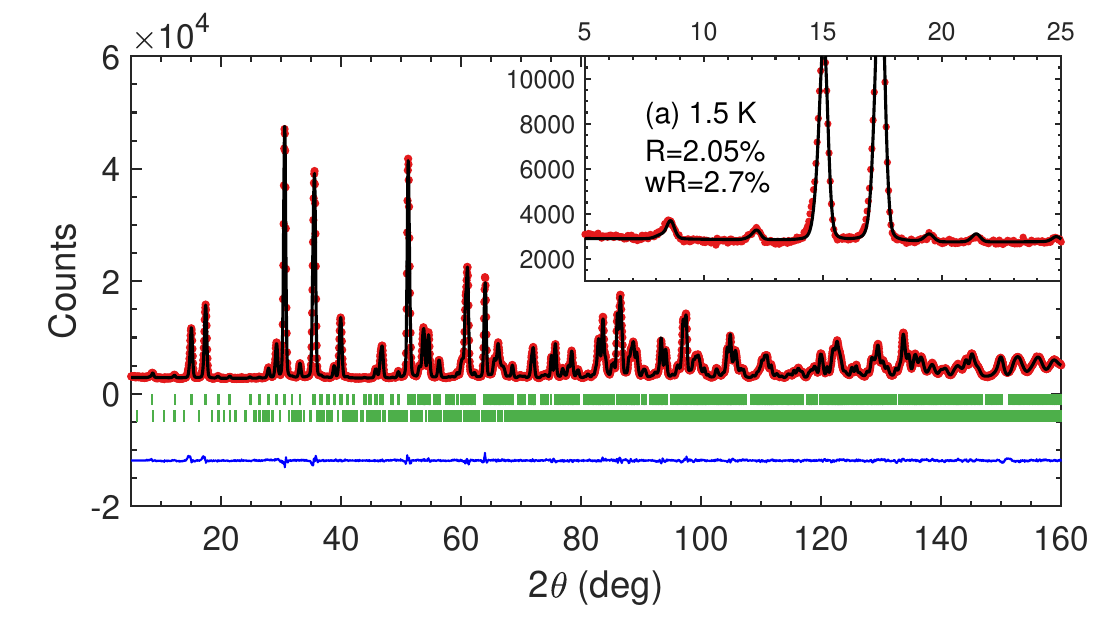}
	\includegraphics[width=0.95\columnwidth]{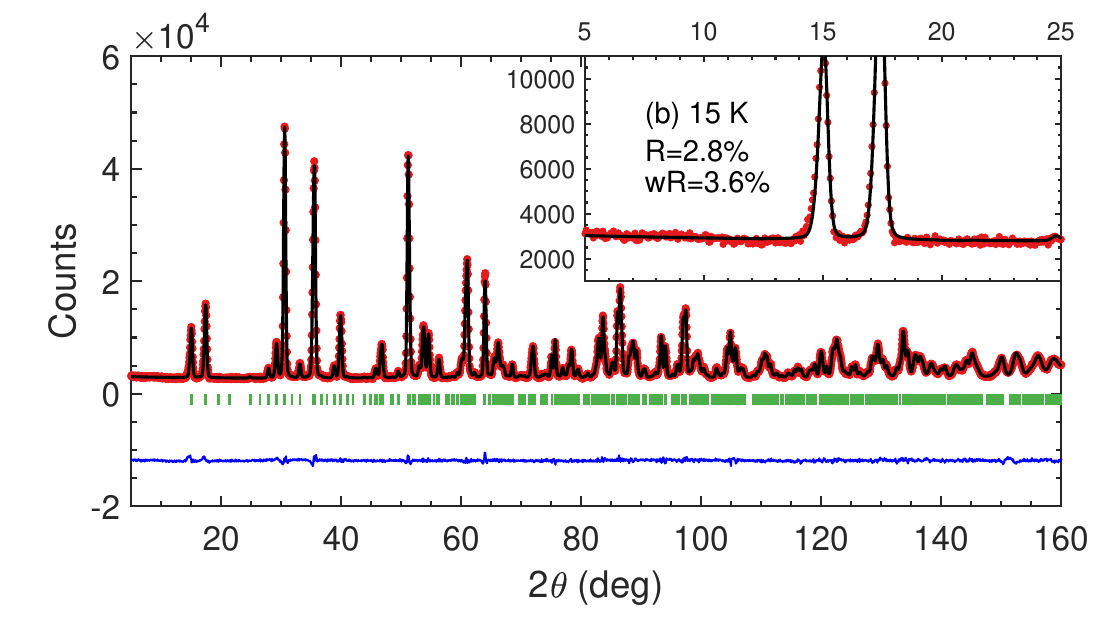}
	\caption{\label{HRPT_patterns} High-resolution PND patterns measured on \krc , (a)
		in the triclinic antiferromagnetic phase at $T=1.5$~K and (b)  at $T=15~K$ in the monoclinic $P2_1/n$ phase. The red circles indicate the experimental data points, the solid black lines illustrate the Rietveld refinements with monoclinic space group $P2_1/n$ (b) , and with both a triclinic nuclear phase in space group $P\bar{1}$ and a magnetic phase described in $P_S\bar{1}$ (a). The corresponding Bragg-peak positions are indicated by the green markers. The blue curve denotes the difference between the fit and the experimental data. In the panels we give the $R$ values corrected for background and the insets present a zoom on the low-angle part, where magnetic
		Bragg peaks are most prominent.}
\end{figure}	

\subsection{Triclinic structural phase in the antiferromagnetic state}
\label{KRC_triclinic}

K$_2$ReCl$_6$ undergoes a magnetic transition at $T_{\rm N}=12$\,K to a long-range AFM order described by the triclinic magnetic subgroup $P_S\bar{1}$~\cite{Bertin24a}. 
The small discontinuity observed at $T_{\rm N}$ in a thermal expansion measurement by dilatometry is a hint for a symmetry-lowering structural phase transition expected to occur at the onset of magnetic order~\cite{Bertin24a}.

A Rietveld analysis with a reference data set measured at 15\,K is shown in Fig.~\ref{HRPT_patterns} and the refined structural parameters are given in Tab. I. The insets in Fig.~\ref{HRPT_patterns} present the emergence of the magnetic Bragg peaks that can be easily resolved with these
high-resolution data.
The magnetic structure is described with the magnetic propagation vector ${\bm k}_{\rm mag}$=($\frac{1}{2}$,$\frac{1}{2}$,0) with regards to the monoclinic unit cell $P2_1/n$~\cite{Bertin24a}. Following the group-subgroup analysis, the maximum magnetic subgroup is $P_S\bar{1}$ with the following unit cell transformation ${\bm a}_{tri}=-{\bm a}_m+{\bm b}_m-{\bm c}_m$, ${\bm b}_{tri}=-{\bm a}_m+{\bm b}_m$, and ${\bm c}_{tri}=2{\bm b}_m$~\cite{isotropysuite}. The breaking of the structural symmetry elements $2_1$ or $n$
splits the atomic positions and thus induces a doubling of the number of positional parameters.
Refinements with this model converge but do not yield any significant differences
between these split sites due to the strongly enhanced number of parameters.
Therefore, we described the nuclear structure by adding 
the corresponding constraints for the position parameters of sites related by symmetry in $P2_1/n$. 
Note, that in this refinement the octahedron can still be distorted as in the paramagnetic $P2_1/n$ phase just above the
N\'eel temperature. 
We also use the same primitive
lattice of  $P2_1/n$ and just allow $\alpha_{\rm t}$ and $\gamma_{\rm t}$ to deviate from
90$^\circ$. 
As can be seen in Tab. I, there is almost no significant change in the positional parameters
at 15 and 1.5\,K, but the triclinic angles clearly deviate from 90$^\circ$ in the antiferromagnetic
phase and there are small but significant changes in the lattice constants; see Fig. 5. The effect in $\gamma$ is 
most pronounced, which perfectly agrees with the magnetic structure model requiring that [0.5,0.5,0] and
[-0.5,0.5,0] become inequivalent due to the emergence of antiferromagnetic order~\cite{Bertin24a}.
Also the moments shown in Fig.~5(c) agree with the previous determination of the magnetic structure that was performed with
higher statistics and a longer wavelength~\cite{Bertin24a}. The length of the moment can be precisely determined
to 1.92(9)\,$\mu_{\mathrm B}$ (compared to 1.98(8)\,$\mu_{\mathrm B}$ in reference \cite{Bertin24a}), but the estimation rotation of the moment with the $a$,$b$ plane suffers
from the small difference between these two monoclinic axes.

\begin{figure}[h!]
	\centering
	\includegraphics[width=1.0\columnwidth]{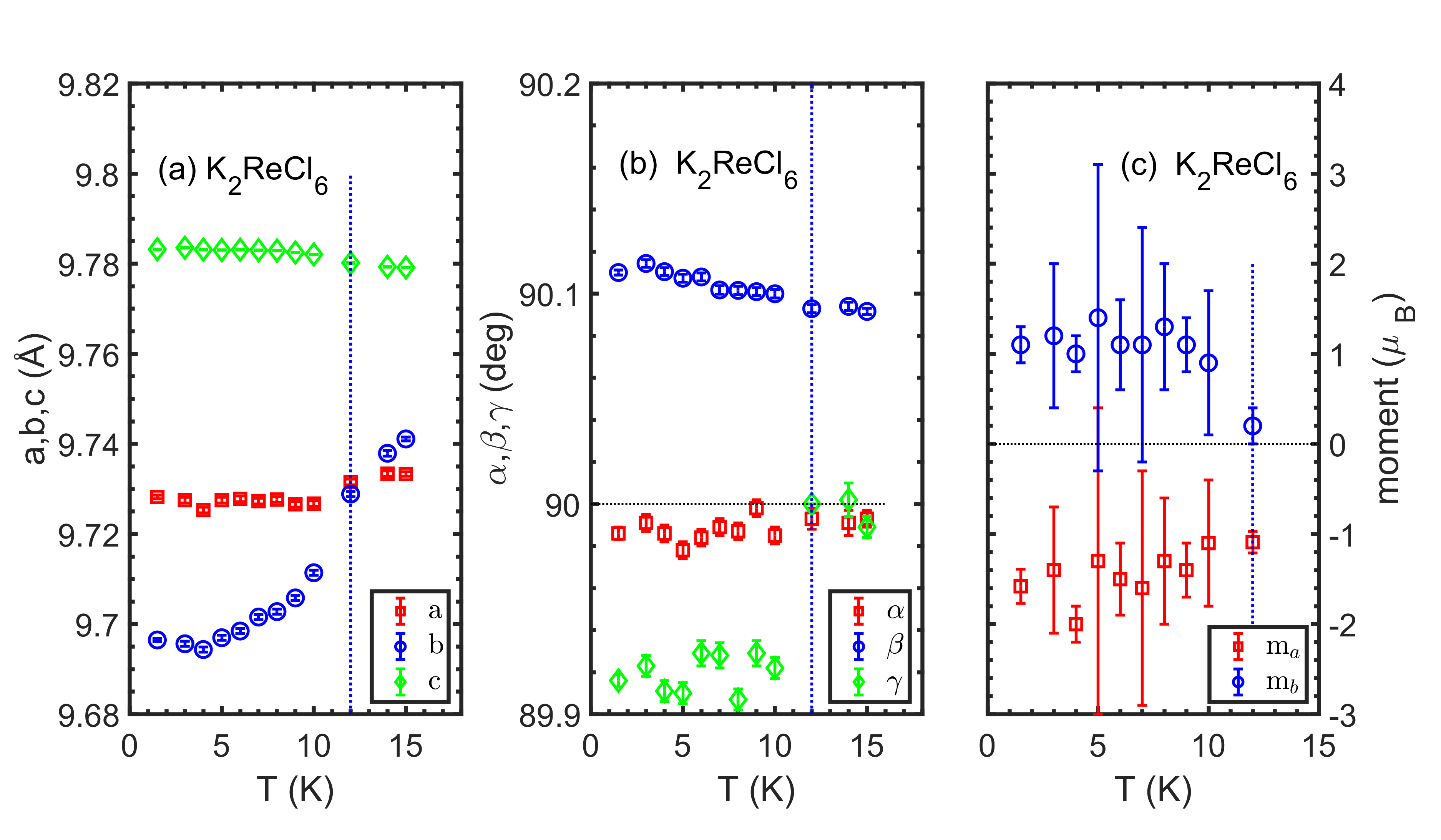}
	
	\caption{\label{HRPT_Tdep_lat} Temperature dependencies of the triclinic lattice parameters $a$, $b$ and  $c$  (a), of the triclinic angles $\alpha$, $\beta$ and $\gamma$ (b)  and of the ordered magnetic moments along the (almost perpendicular) directions $a$ and $b$ (c) that were deduced from Rietveld refinements with high-resolution PND patterns on HRPT. The magnetic transition at $T_{\rm N}=12$\,K is indicated by the vertical blue dotted line.}
\end{figure}

\subsection{Crystal structure of \ksc ~ as a reference material with completely filled $d$ shell}

K$_2$SnCl$_6$ crystallizes at room temperature in the cubic antifluorite structure $Fm\bar{3}m$, and exhibits a sequence of structural transitions in a narrow temperature range. First, two structural transitions have been revealed by means of powder and single-crystal neutron and x-ray diffraction experiments at 261\,K towards a tetragonal phase described with space group $P4/mnc$, and at 255\,K towards a monoclinic phase with space group $P2_1/n$~\cite{Boysen76,Boysen78}. These structural transitions support the interpretation of Raman scattering and M\"o\ss bauer spectra~\cite{Winter76,Pelzl77}. NQR measurements also suggest two structural transitions at 262\,K and 256\,K, but towards first an orthorhombic structure and then to a monoclinic phase~\cite{Seo98}. Investigation of x-ray diffuse scattering at room temperature suggests that the first structural transition is driven by the condensation of a soft phonon mode at the $\Gamma$ point~\cite{Ihringer80}, similarly to early studies on K$_2$ReCl$_6$~\cite{Lynn78,Armstrong80}. However, later, the temperature study of selected Bragg reflections by high-resolution SXD in reference \cite{Kugler83} revealed the onset of an intermediate structural phase at 265\,K, before a clear orthorhombic splitting is evidenced at 262\,K, and the monoclinic structure sets in at 255\,K. 
In reference~\cite{Ihringer84} the structural sequence was then further clarified: three structural transitions occur at $T_{\rm t}=263$\,K, $T_{\rm m1}=261$\,K, and $T_{\rm m2}=255$\,K towards a tetragonal phase with space group $P4/mnc$, a first monoclinic phase with space group $C2/c$, and a second monoclinic phase with space group $P2_1/n$. Therefore, in a narrow temperature range, K$_2$SnCl$_6$ exhibits the same sequence of structural phase transitions as K$_2$ReCl$_6$. Due to its fully occupied $4d^{10}$ electronic configuration, the Jahn-Teller effect is not active rendering K$_2$SnCl$_6$ an ideal reference material.

In order to compare with our results on K$_2$ReCl$_6$, SXD data have been collected in the four different crystallographic phases: at 270\,K in the cubic $Fm\bar{3}m$ phase, at 262\,K in the tetragonal $P4/mnc$ phase, at 261\,K, 259\,K, and 257\,K in the first $C2/c$ monoclinic phase, and at 230\,K, 200\,K, 150\,K, 95\,K, and 82\,K in the second monoclinic $P2_1/n$ phase. The data completeness, and the $R$-values and GOF parameters documenting the quality of the refinements are tabulated in Tab.~S1 in the supplemental material \cite{supplmat}.

\begin{figure}[h!]
	\centering
	\includegraphics[width=0.99\columnwidth]{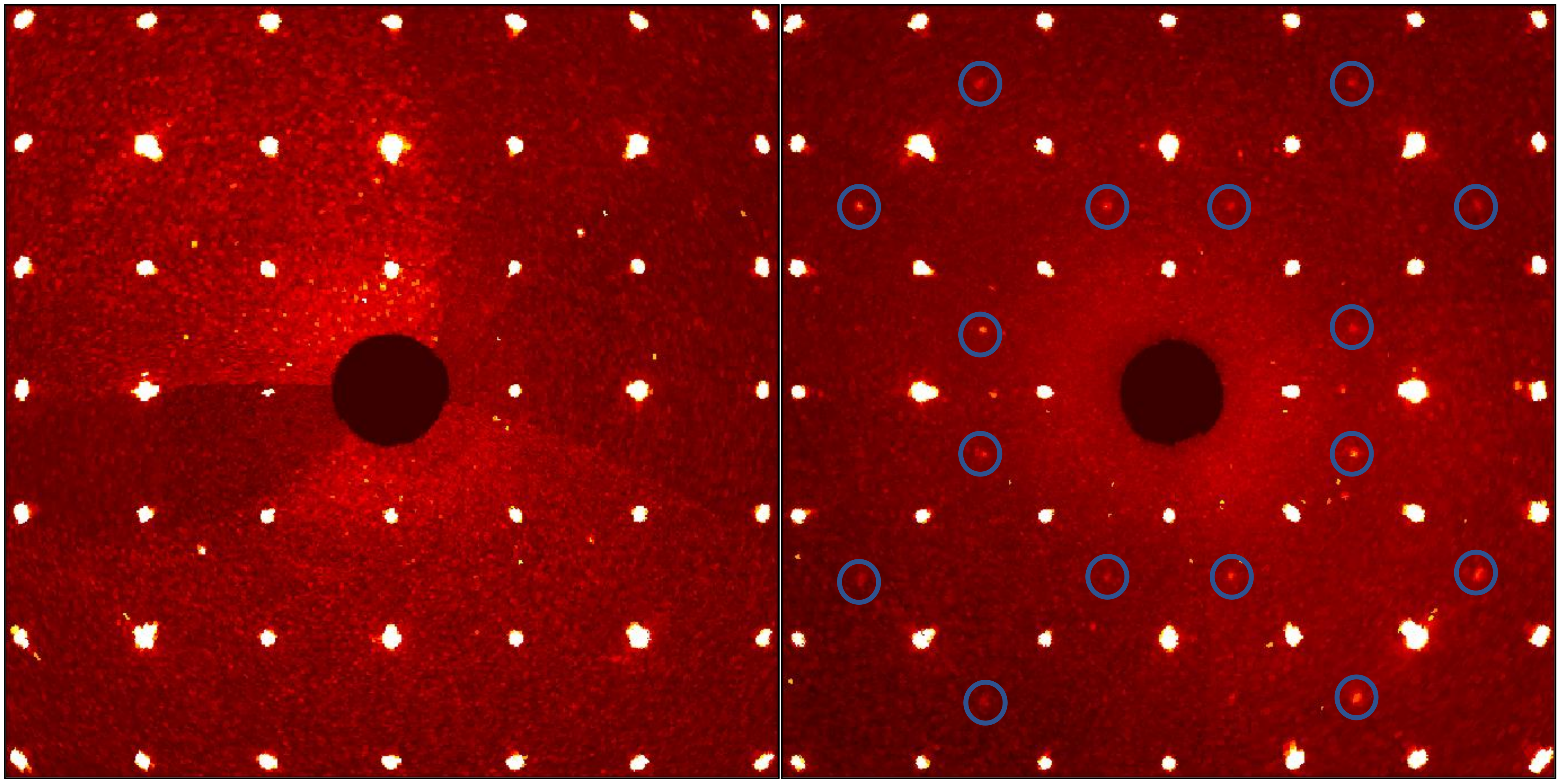}	
	\caption{\label{KSC_maps}  Precession maps computed in the $(h0l)$ scattering plane from SXD measurements on K$_2$SnCl$_6$ at 270\,K in the cubic phase (left panel) and at 262\,K in the tetragonal phase (right panel). The blue circles highlight the superstructure reflections emerging in the low-symmetry phase, and breaking the $F$ centering. }
\end{figure}

Precession maps calculated in the $(h0l)$ reciprocal space plane are shown in Fig.~\ref{KSC_maps} at 270\,K in the cubic phase (left panel) showing the good quality of our crystal, and at 262\,K in the tetragonal phase (right panel). Clear superstructure reflections of type (103), (301), (105) and (503) indicate the breaking of the $F$ centering and attest the presence of an intermediate tetragonal phase in a narrow temperature range, supporting the conclusions from Ref.~\cite{Ihringer84}. Note also the presence of minor spurious reflections arising from a tiny additional grain.

Data sets collected in the low-symmetry phases were treated as described in Sec.~II. All inferred structural information are gathered in Tab.~S3 in the supplemental material \cite{supplmat}. Note that in the two monoclinic phases, the structural model solutions describing best the data sets are not unique. This ambiguity is lifted by the Rietveld analysis of high-resolution PND data collected in both monoclinic phases: the chosen structural model solution corresponds to the choice of monoclinic angle $\beta_{\rm m1,m2}>90^{\circ}$. At 262\,K, in the tetragonal phase, there is a strong correlation between $x$(Cl$_1$) and $z$(Cl$_3$) that cannot be suppressed by fixing $U_{\parallel}$(Cl) (to a value linearly scaled from the $U_{\parallel}$(Cl) value found at 270\,K), as it was done in the tetragonal phase of K$_2$ReCl$_6$. Note that in all subsequent refinements, no other constraints or fixed parameters are introduced, and all parameters are refined, including all twining fractions.

Since the sequence of structural phase transitions is identical to that in K$_2$ReCl$_6$, we can directly compare the octahedral rotation/tilt angles $\phi/\theta$ in Fig.~\ref{KRC_KSC_all}(b), which includes the PND results. The increase of $\phi$ is consistent with a second-order phase transition, while the tilt angle $\theta$ exhibits a discontinuous increase at the first monoclinic structural transition $T_{\rm m2}=260$\,K with $\theta=3.6(2)^{\circ}$ suggesting a first-order character. Note that already in the tetragonal phase, the rotation angle $\phi=2.23(3)^{\circ}$ is larger than that found in the tetragonal phase of K$_2$ReCl$_6$, explaining the stronger superstructure reflections observed in the precession maps. The tilt and rotation angles increase continuously up to $\phi =\theta = 8.1(1)^{\circ}$, a value also roughly twice larger than those in K$_2$ReCl$_6$, in line with the smaller tolerance factor calculated for K$_2$SnCl$_6$ due to the larger radius of the Sn$^{4+}$ ion~\cite{Bertin24b}.

Figure~\ref{KRC_KSC_all} presents the temperature dependence of the ADPs in \ksc .
$U_{\perp}$(Cl) and $U_{\rm iso}$(K) strongly shrink across the narrow temperature interval of the three structural transitions, which reflects the more abrupt emergence of the tilt and rotations distortions and points to some disorder in the cubic phase; see below.
In the low-temperature phase $P2_1/n$, all ADPs follow the expected roughly linear behavior upon cooling.

For the \ksc ~ SXD measurement at 270\,K in the cubic phase,
the anharmonic refinement with space group $Fm\bar{3}m$ yields considerable improvement of the $R$-values; see Tab. S5 in the supplemnental material \cite{supplmat}, and the most significant anharmonic parameter $C_{122}=C_{133}$ is much larger than that found in the Re counterpart.
Furthermore, the split model yields a much stronger deviation from the cubic symmetry $y({\rm Cl})=0.0206(7)$, compared to K$_2$ReCl$_6$; see Tab. S5. This deviation of $\approx 0.2$\AA~is comparable to $\sqrt{U_{\perp}}=0.28$\AA~deduced from the refinement with the harmonic ADPs. 
In addition, $U_{\perp}$ is strongly reduced in the split model. Finally, there is a significant improvement of the $R$-values compared to the standard analysis with harmonic ADPs.
For K$_2$SnCl$_6$, both methods thus indicate the presence of some disorder at the Cl site in the cubic phase
that may stem from a small deviation from stoichiometry; see section III.A,
but note that the measurement was performed more close to the structural phase transition than in the case of \krc . 
Local disorder in the parent phase furthermore can contribute to the more rapid emergence of the long-range rotational distortion upon cooling that is visible in Fig.~2 and \ref{KSC_HRPT_lat_param} as well as in the literature \cite{Kugler83}. The low ADPs of the Cl sites in \ksc ~ at low temperature, however, clearly show that the low-temperature phase is well ordered and does not exhibit a significant local rotational disorder. Note, that a fully disordered rotation
by about 8$^\circ$ would imply   $U_{\perp} \sim 0.1$\,\AA$^2$ while the low-temperature
experimental value is an order of magnitude smaller and consistent with dynamical
zero-point fluctuations (that are thus much smaller than the static distortion).

\begin{figure}[h!]
	\centering
	\includegraphics[width=0.95\columnwidth]{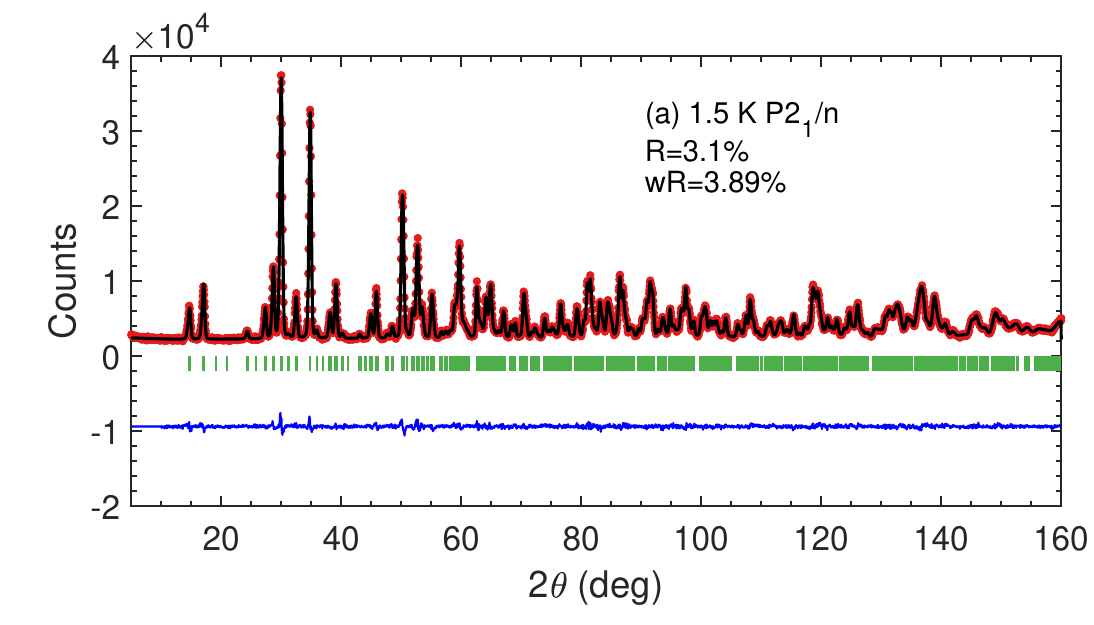}
\includegraphics[width=0.95\columnwidth]{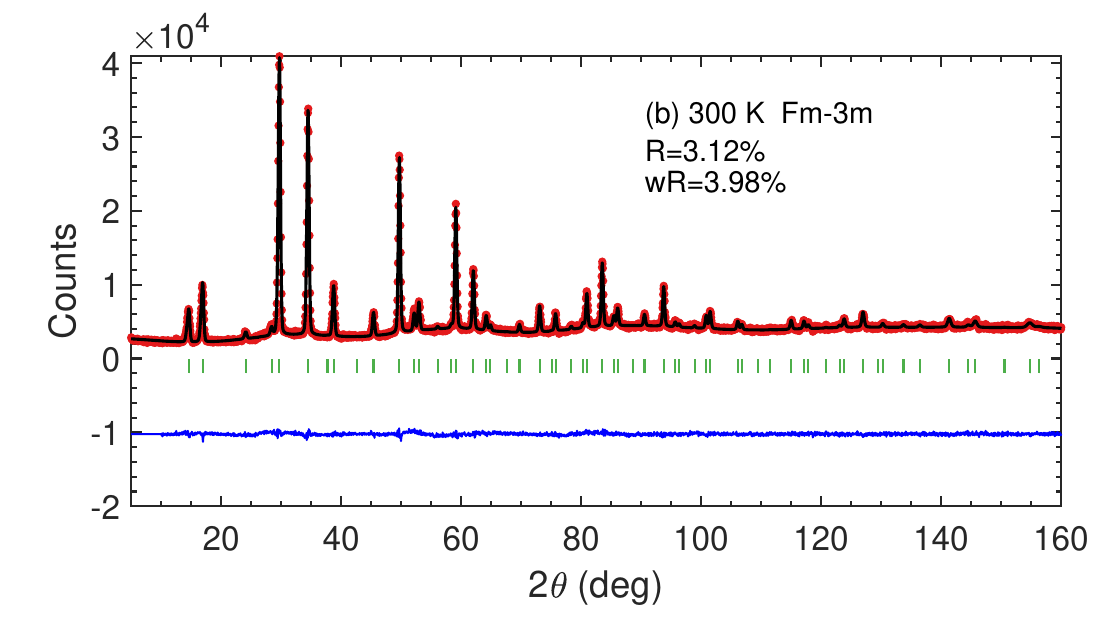}
	\caption{\label{HRPT_KSC_patterns} Rietveld analysis of PND patterns measured on \ksc ~ (a) at $T=1.5$~K in the monoclinic $P2_1/n$ phase  and (b) at $T=300$~K in the cubic $Fm\bar{3}m$ phase (reference pattern) . The signification of colored lines and bars corresponds to Fig.~3. }
\end{figure}

The temperature dependencies of the lattice parameters and of the monoclinic angles, deduced from the PND Rietveld analysis, are shown in Fig.~\ref{KSC_HRPT_lat_param}, which also presents a zoom into the narrow temperature range where the three structural transitions occur. A clear tetragonal splitting is resolved at $T^{\prime}_{\rm t}=260$\,K indicating the transition to a tetragonal phase with space group $P4/mnc$, followed by breaking of the tetragonal symmetry and the deviation from 90$^{\circ}$ of the monoclinic angle $\beta_{m1}$ at $T^{\prime}_{\rm m1}=257$\,K marking the transition to the first monoclinic phase described with the $C2/c$ space group. Finally, the transition towards the second monoclinic phase with space group $P2_1/n$ occurs below  $T^{\prime}_{\rm m2}<252$\,K. The sequence of phase transitions in our sample thus agrees with reference~\cite{Ihringer84}. 

The tetragonal splitting and the monoclinic angles are larger in \ksc ~ in agreement with the larger tilt and rotation angels. A crossing between the $a_{\rm m2}$ and $b_{\rm m2}$ happened around 120\,K and was not evidenced in K$_2$ReCl$_6$. 
The correct assignment of the lattice parameters has been carefully tested as it is explained in the supplemental material \cite{supplmat}.

\begin{figure}[h!]
	\centering
	\includegraphics[width=0.99999\columnwidth]{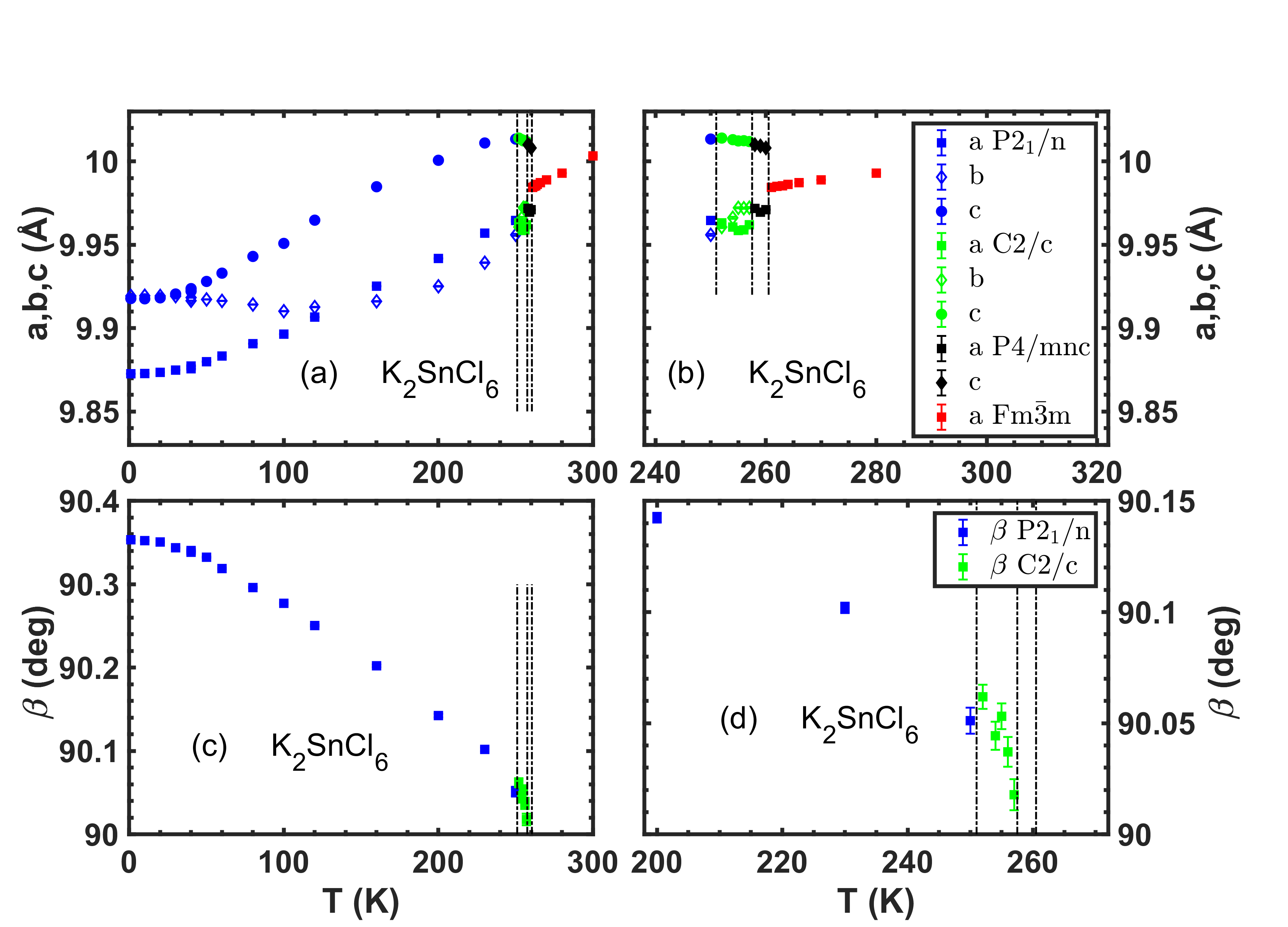}	
	\caption{\label{KSC_HRPT_lat_param} Temperature dependence of the K$_2$SnCl$_6$ lattice parameters inferred from Rietveld refinements describing the PND patterns measured on HRPT. A zoom across the successive structural transitions over a narrow temperature range is shown in the right panels. The black dashed lines indicated the structural transition temperatures from our neutron powder measurements, i.e.\ $T^{\prime}$ not corrected for the temperature offset $\Delta T$.}
\end{figure}

A pattern measured at 1.5\,K in the monoclinic phase $P2_1/n$ is shown in Fig.~\ref{HRPT_KSC_patterns} and can be compared with the pattern measured at room temperature. These patterns appear quite different due to the strong structural distortions
reducing the lattice symmetry and due to the large ADPs reducing the high-angle intensities in the
300\,K pattern. 
As the high-angle part of these data further suffers from the lower resolution in these PND experiments compared to
that on \krc ~ the precision of the structural parameters at higher temperature is no longer comparable to that in
the SXD experiments.
Therefore, only the low-temperature Rietveld results for \ksc ~ are included in
Fig.~\ref{KRC_KSC_all} and Fig.~\ref{distortion}(b).

\section{Absence of Jahn-Teller distortions in  {K$_2$R\lowercase{e}C\lowercase{l}$_6$} at low temperature}
\label{discussion}

\begin{figure}[htb!]
	\centering
	\includegraphics[width=0.999\columnwidth]{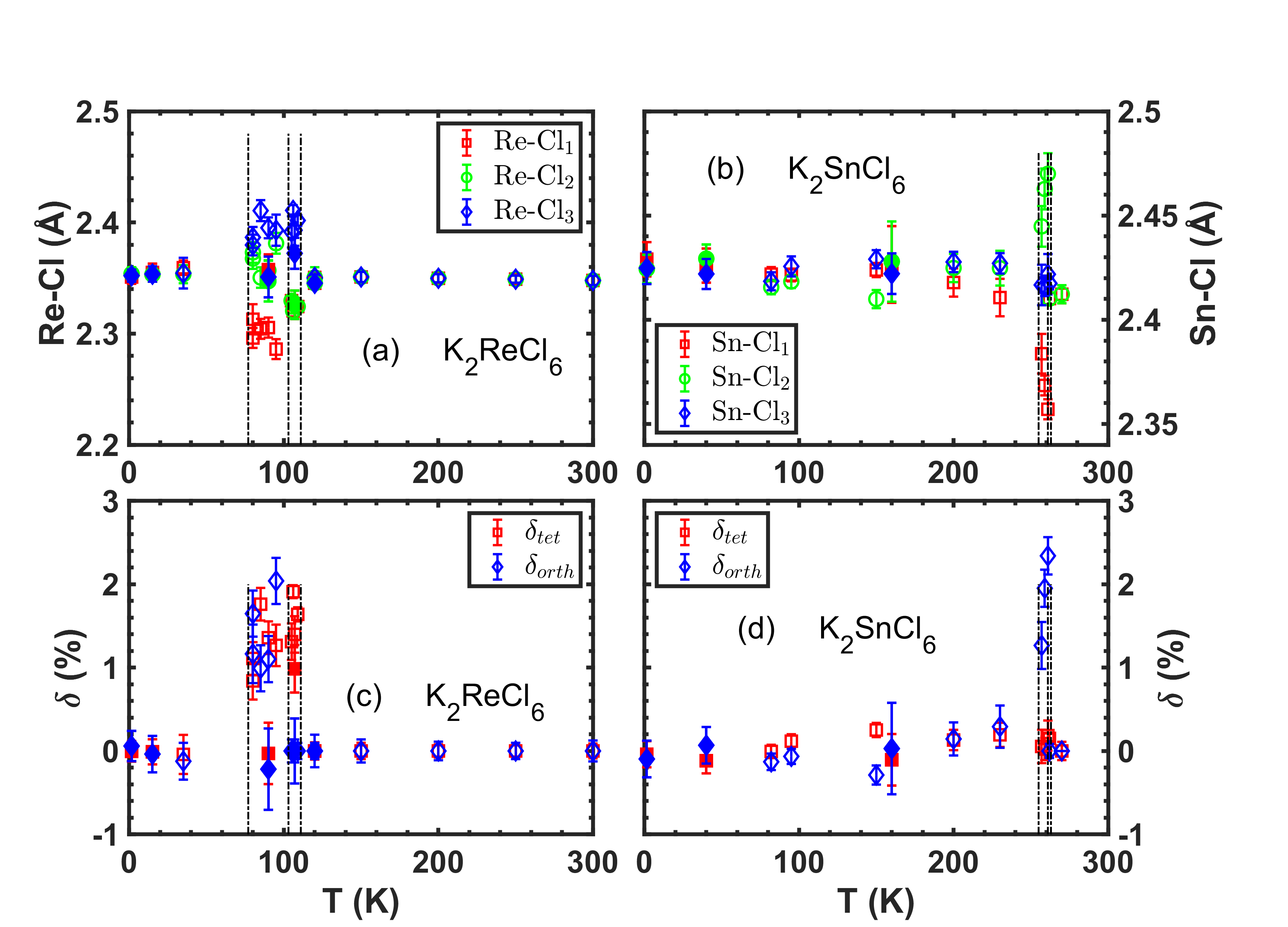}
	\caption{\label{distortion} Temperature dependencies of the $Me$-Cl distances (a,b) and of the averaged tetragonal and orthorhombic octahedral distortions (c,d) as defined in the main text, for \krc \ in panel (a,c)
		and for \krc ~ in (b,d). Open symbols refer to the K$_2$ReCl$_6$ SXD and to the \ksc \ SXD results and the filled symbols designate the PND analyses. Vertical dashed black lines indicate the structural transitions.}
\end{figure}

In order to analyze the possible presence of a Jahn-Teller distortion, the tetragonal, $\delta_{tet}$, and orthorhombic, $\delta_{orth}$, octahedral distortions are calculated as follows:
\begin{equation}
	\delta_{tet}= \frac{d_3-d_{bas}}{d_3+d_{bas}};~~~\delta_{orth}=\frac{d_2-d_1}{d_2+d_1},
\end{equation}
where $d_i$ are the Re-Cl$_i$ bond distances, and $d_{bas}=(d_1+d_2)/2$ is the average distance in the basal plane. While $d_{bas}=d_1=d_2$ in the tetragonal phase, $d_1\ne d_2$ in the two monoclinic phases. 
%All bond distances and octahedral rotation/tilt angles are calculated using atomic positions derived from SXD and PND data in combination with lattice parameters obtained from powder x-ray diffraction~\cite{Bertin24a} for \krc \ and PND for K$_2$SnCl$_6$. 
The octahedral distortion parameters are plotted against temperature in Fig.~\ref{distortion} showing good agreement between SXD and PND results.

Focusing first on the K$_2$ReCl$_6$ results shown in Fig.~\ref{distortion}(a,c), a strong tetragonal distortion with an average value of $\delta_{tet} \approx 1.6$\% emerges in the tetragonal phase $P4/mnc$, reaches a maximum in the first monoclinic phase $C2/c$ and decreases when approaching the transition to the second monoclinic phase $P2_1/n$. At low temperature in $P2_1/n$, SXD and PND experiments consistently yield a vanishing tetragonal distortion.
A significant orthorhombic distortion $\delta_{orth} \approx 1.3$\% emerges in the upper monoclinic phase $C2/c$, but again no sizable distortion persists at low temperature in the $P2_1/n$ phase.
The presence of octahedral distortions at intermediate temperatures and their disappearance at the lowest temperatures is also visible in the plot of the three Re-Cl distances in Fig. 10(a).

Figure~\ref{distortion}(b,d) presents the temperature dependencies of the Sn-Cl distances and of both octahedral distortions in \ksc \ \footnote{Because of the temperature offset $\Delta$T in the PND measurements, in order to calculate the bond-distances and octahedral rotation/tilt angles, lattice parameters computed at 260\,K (first refinement with a tetragonal cell) and 254\,K are used in combination with the single crystal refinements at 262\,K ($P4/mnc$) and 257\,K ($C2/c$). At 150\,K, the lattice parameters are averaged from the results obtained at 200\,K and 100\,K.}.
Unlike in K$_2$ReCl$_6$, the tetragonal distortion remains below 0.3\% over the full temperature range. On the contrary, similarly to K$_2$ReCl$_6$, in the intermediate monoclinic $C2/c$ phase, a large orthorhombic distortion $\delta_{orth}=2.3(1)$\% emerges at $T_{\rm m1}$, and decreases towards low temperature. Upon entering in the second $P2_1/n$ monoclinic phase, the orthorhombic distortion becomes vanishingly small down to the lowest studied temperature.

In the same spirit one may analyze the Cl-Cl distances corresponding to the octahedral edges, see supplemental material~\cite{supplmat}.
In the cubic phase all edges are equivalent with a length of 3.34\,\AA ~ in \krc ~ (3.41\,\AA ~ in \ksc).
In the low temperature phases there can be up to six different edge lengths and at intermediate temperature
one observes several splittings of the order of one percent, but at low temperature all edge lengths are
nearly identical again. The absence of sizable splitting of Cl-Cl and Re-Cl distances implies that also the bond-angles in an octahedron
do not deviate from 90$^\circ$.
 
Summarizing the analysis of octahedral distortions, K$_2$SnCl$_6$ exhibits a sizable orthorhombic octahedral distortion in the first monoclinic phase $C2/c$ that quickly vanishes when entering the second monoclinic $P2_1/n$ phase. No tetragonal distortion is perceptible in \ksc \ across the entire temperature range.
K$_2$ReCl$_6$ also exhibits sizable orthorhombic and tetragonal octahedral distortions but only in the intermediate structural phases, where it thus differs from the reference compound \ksc .
However, in the low-temperature  $P2_1/n$ phase, both octahedral distortions relax and nearly vanish also in \krc . This renders an electronic or Jahn-Teller origin of the distortions very unlikely in agreement with RIXS experiments on \krc ~\cite{Warzanowski24}.  The distortions appearing in the intermediate phases of both materials seem to stem from a purely structural origin.
In the tetragonal and upper monoclinic phases the rotated octahedron cannot fully relax but seems
to be forced into these distortions by structural constraints, which however finally disappear with the other tilting scheme in the low-temperature symmetry $P2_1/n$. The suppression of internal octahedral distortions
in the low-temperature $P2_1/n$ phase contributes to the balance between the two monoclinic phases that both combine tilt and rotations. Therefore the relaxation of octahedral deformations contributes to driving this first-order phase transition in the two materials with completely and partially filled $d$ shells.

\section{Magnetic order and crystal structure in {R\lowercase{b}$_2$R\lowercase{e}C\lowercase{l}$_6$} and {C\lowercase{s}$_2$R\lowercase{e}C\lowercase{l}$_6$} }

It appears interesting to investigate the magnetic order and crystal structure in the related materials
Rb$_2$ReCl$_6$ and Cs$_2$ReCl$_6$, because the larger alkaline ionic radii of Rb$^+$ and Cs$^+$ imply larger tolerance factors of 1.023 and 1.069, respectively. For these tolerance-factor values, one does not expect the rotational phase transitions \cite{Bertin24b}, while the 
electronic configuration of Re$^{4+}$ is identical to that in \krc . These two materials 
were very little studied: For Cs$_2$ReCl$_6$ the room temperature crystal structure in space group $Fm\bar{3}m$ and magnetic measurements above 90\,K were reported \cite{Sperka1988,Figgis1961} and for Rb$_2$ReCl$_6$ only luminiscence experiments were published \cite{Bettinelli1988,Bettinelli1991}.

\begin{table}[h!]
	\centering
	\caption{\label{RbCsReCl} Magnetic properties and crystal structure parameters of  Rb$_2$ReCl$_6$ and Cs$_2$ReCl$_6$. The analysis of the SQUID magnetization measurements at 0.1\,T yields the magnetic
		entities in the upper part, and Rietveld refinements in space group $Fm\bar{3}m$ with powder x-ray diffraction data are summarized in the lower part. Re is at (0,0,0) and Rb/Cs at ($\frac{1}{4}$,$\frac{1}{4}$,$\frac{1}{4}$) and atomic displacement parameters $U$  are given in \AA $^2$. }
	\begin{tabular}{c|cc|cc}
		\hline
		& \multicolumn{2}{c|}{Rb$_2$ReCl$_6$}   &\multicolumn{2}{c}{Cs$_2$ReCl$_6$} \cr
		\hline
		$\chi_0$ (emu/mol) & \multicolumn{2}{c|}{-9.4(2)$\cdot$10$^{-4}$} & \multicolumn{2}{c}{-8.49(8)$\cdot$10$^{-4}$} \cr
		$\theta$ (K) & \multicolumn{2}{c|}{-92.7(5)} & \multicolumn{2}{c}{-45.5(5)} \cr
		$p_{\mathrm{eff}}$ ($\mu_{\mathrm B}$) & \multicolumn{2}{c|}{3.93(2)} & \multicolumn{2}{c}{3.79(1)} \cr
		$T_\mathrm{N}$ (K) & \multicolumn{2}{c|}{15.0(6)} & \multicolumn{2}{c}{9.4(6)} \cr
		\hline
		$T$ (K)  &    290 &   5 & 290 & 7 \cr		
		$R$ (\%) &    7.5 & 10.1 & 7.4 & 7.9 \cr
		$a$ (\AA )  &    9.9738(2) &   9.8684(2) & 10.2578(2) & 10.1510(2) \cr		
		Re  $U_{iso}$ & 0.0378(4) & 0.0169(4) & 0.0242(5) & 0.0108(4) \cr
		Rb/Cs   $U_{iso}$ &  0.0482(6) & 0.0131(5) & 0.0331(6) & 0.0141(4) \cr
		Cl ($x$,0,0) & 0.2462(2) & 0.2489(3) & 0.2425(3) & 0.2444(3) \cr
		~~$U_{11}$ & 0.036(2) & 0.004(2) & 0.019(2) & 0.012(2) \cr
		~~$U_{22}$=$U_{33}$ & 0.069(2) & 0.030(2) & 0.048(2) & 0.028(2) \cr
		\hline			
	\end{tabular}
\end{table}

\begin{figure}[htb!]
	\centering
	\includegraphics[width=0.99\columnwidth]{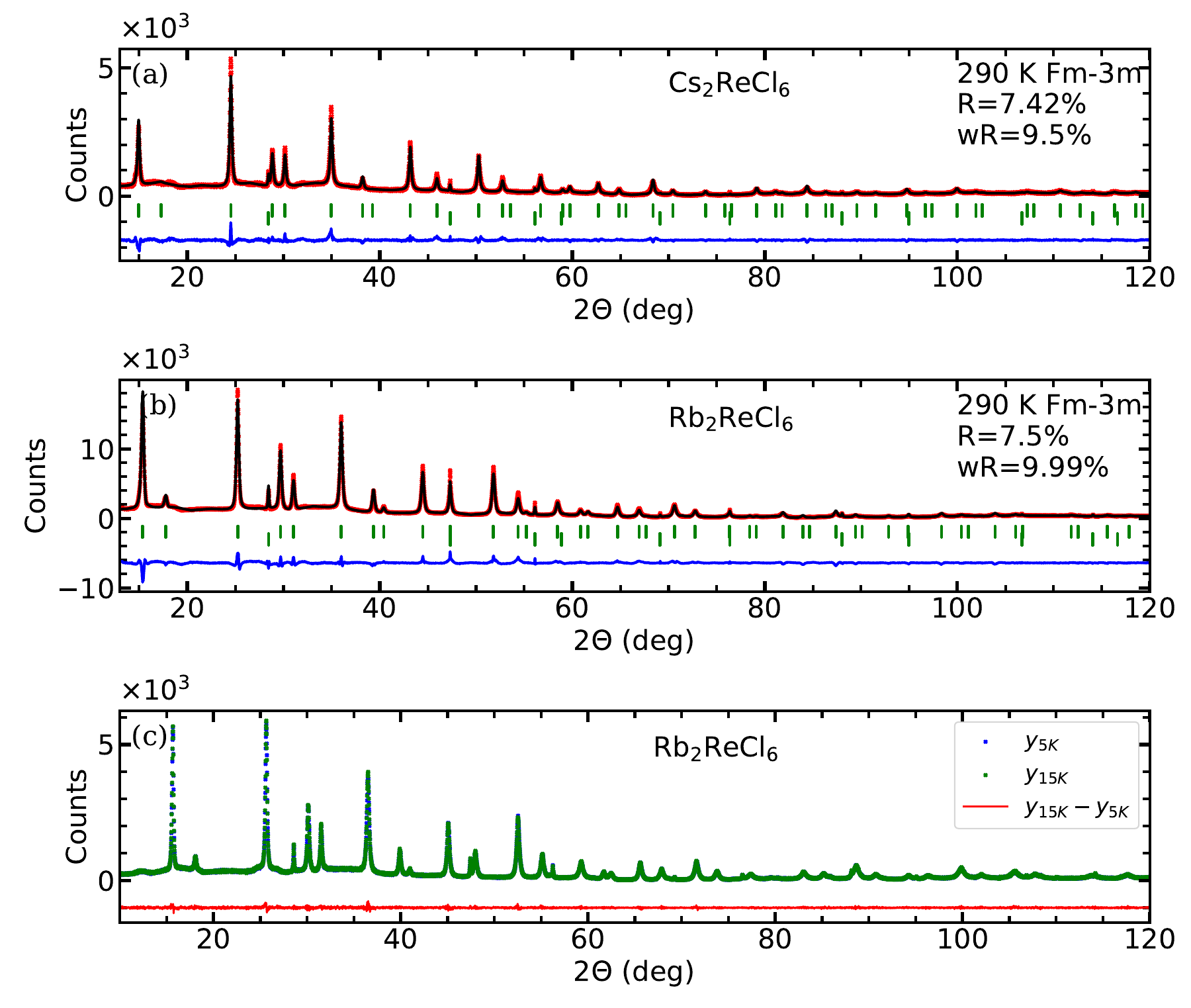}
	\caption{\label{xrayStoe} Rietveld analysis of the powder x-ray diffraction patterns obtained for  Cs$_2$ReCl$_6$ (a) and Rb$_2$ReCl$_6$ (b) at 290\,K. The signification of colored lines and bars corresponds to Fig. 3. There are no peaks that cannot be indexed with the main phase [besides those of
		the admixed Si powder (lower row of green bars)]. Panel (c) compares the patterns obtained for Rb$_2$ReCl$_6$ at and below its N\'eel temperature of 15\,K as well as the subtraction of the two 
		patterns (red lines).}
\end{figure}

\begin{figure}[htb!]
	\centering
	\includegraphics[width=0.95\columnwidth]{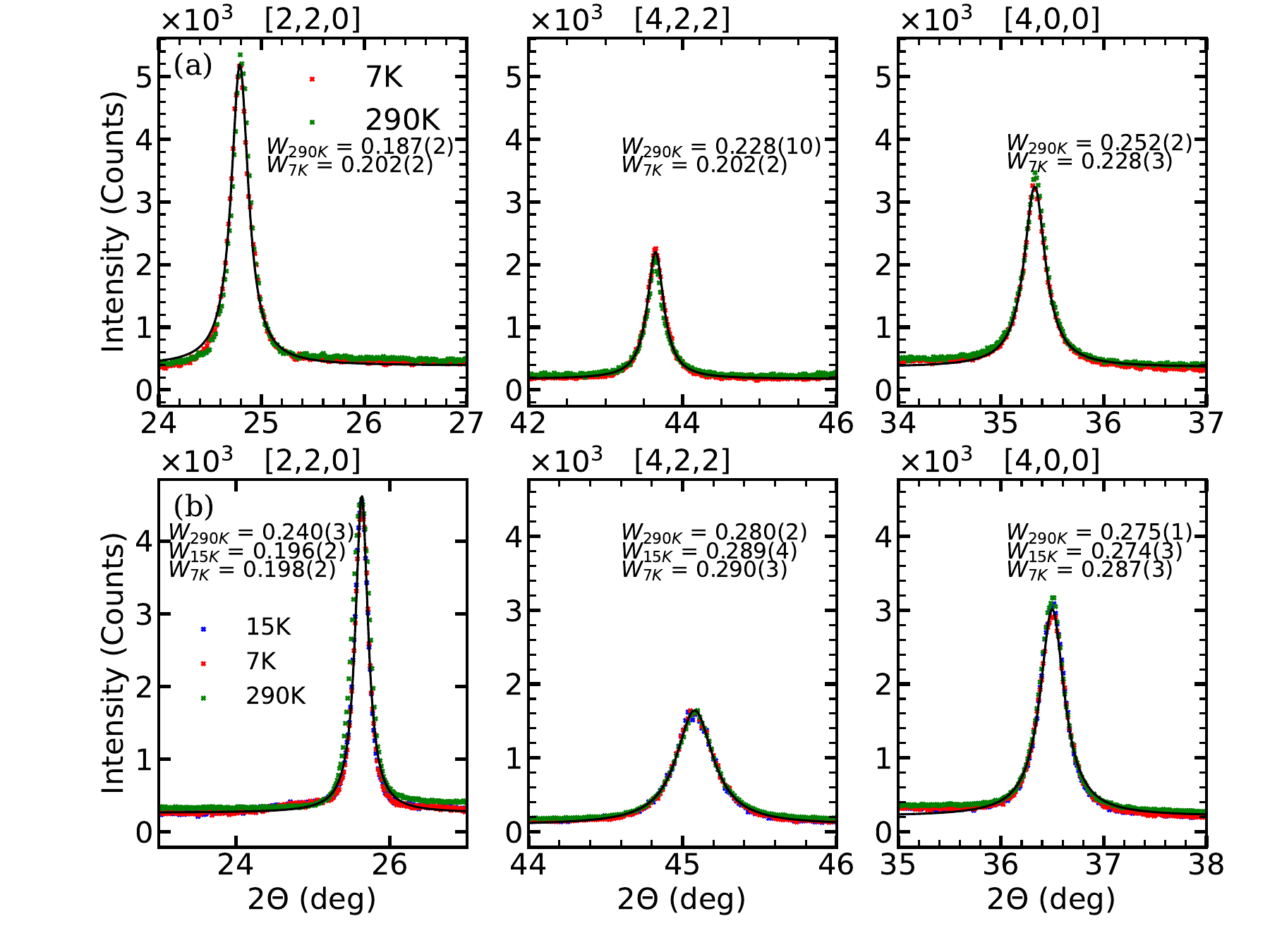}
	\caption{\label{comppeaks} Comparison of several Bragg reflection peak profiles for  Cs$_2$ReCl$_6$ [upper panels (a)] and for Rb$_2$ReCl$_6$ [lower panels (b)]. For   Rb$_2$ReCl$_6$ we compare 7, 15 and 290\,K and for
		Cs$_2$ReCl$_6$ 7 and 290\,K.  We have horizontally shifted the low-temperature data to
		better visualize any changes in the peak profiles. The individual peaks were fitted by Lorentzian
		distributions (black lines for the low-temperature data) and the differences of the width parameters are shown in the panels, but there is no significant change. }
\end{figure}

The powder
x-ray diffraction data obtained with both samples at temperatures of 290\,K and below 10\,K are well described in space group $Fm\bar{3}m$; see Fig.~\ref{xrayStoe} indicating that no structural phase transition occurs below room temperature in these two materials. 
For a closer inspection Fig.~\ref{comppeaks} compares the peak profiles
of several Bragg reflections measured at 290\,K and below 10\,K, respectively. For better visibility the low-temperature data were shifted horizontally to correct for the thermal expansion. There is no change in the peak profiles, and also the fitted widths of Lorentzian distributions do not significantly change upon cooling. Therefore, we can conclude that Rb$_2$ReCl$_6$ and Cs$_2$ReCl$_6$ both do not exhibit the rotational phase transition observed in all antifluorites with a tolerance factor below $\sim$1; see Ref. \cite{Bertin24b}. The absence
of a translation-lattice distortion also renders a long-range Jahn-Teller very unlikely, although it may
not fully be excluded that local octahedral distortions arrange in a scheme maintaining cubic symmetry.  
The powder data do not permit a comparably detailed structural analysis as our measurements on \krc ;
nevertheless the absence of a sizable lattice distortion well agrees with the absence of the Jahn-Teller effect in \krc ~ deduced from our more precise SXD and PND studies. 

The structural parameters obtained from the Rietveld analysis are summarized in Tab. II. There is agreement
with the existing room-temperature data for Cs$_2$ReCl$_6$ \cite{Sperka1988}. 
The room-temperature cubic lattice constants increase with the atomic number and with the size of the alkaline ion. 
All three $A_2$ReCl$_6$ compounds
exhibit a quite large thermal expansion with lattice-constant differences between $\sim$10 and 290\,K  of $\Delta a$=0.107\,\AA ,  $\Delta a$=0.105\,\AA \ and  $\Delta a$=0.096\,\AA \ for the Cs, Rb and K compounds, respectively. The pronounced contraction can be attributed to the empty site
of the antifluorite structure compared to a perovskite, and the smaller volume shrinking in \krc ~
results from the occurrence of the rotational phase transitions that relax internal strain similar to
perovskite related compounds \cite{braden1994}. The Rietveld refinements find large ADPs for the
$A$ site and for the direction perpendicular to the bond at the Cl site, which indicate that the phonon modes associated with rotational transitions are also soft in these two materials.

\begin{figure}[htb!]
	\centering
	\includegraphics[width=0.85\columnwidth]{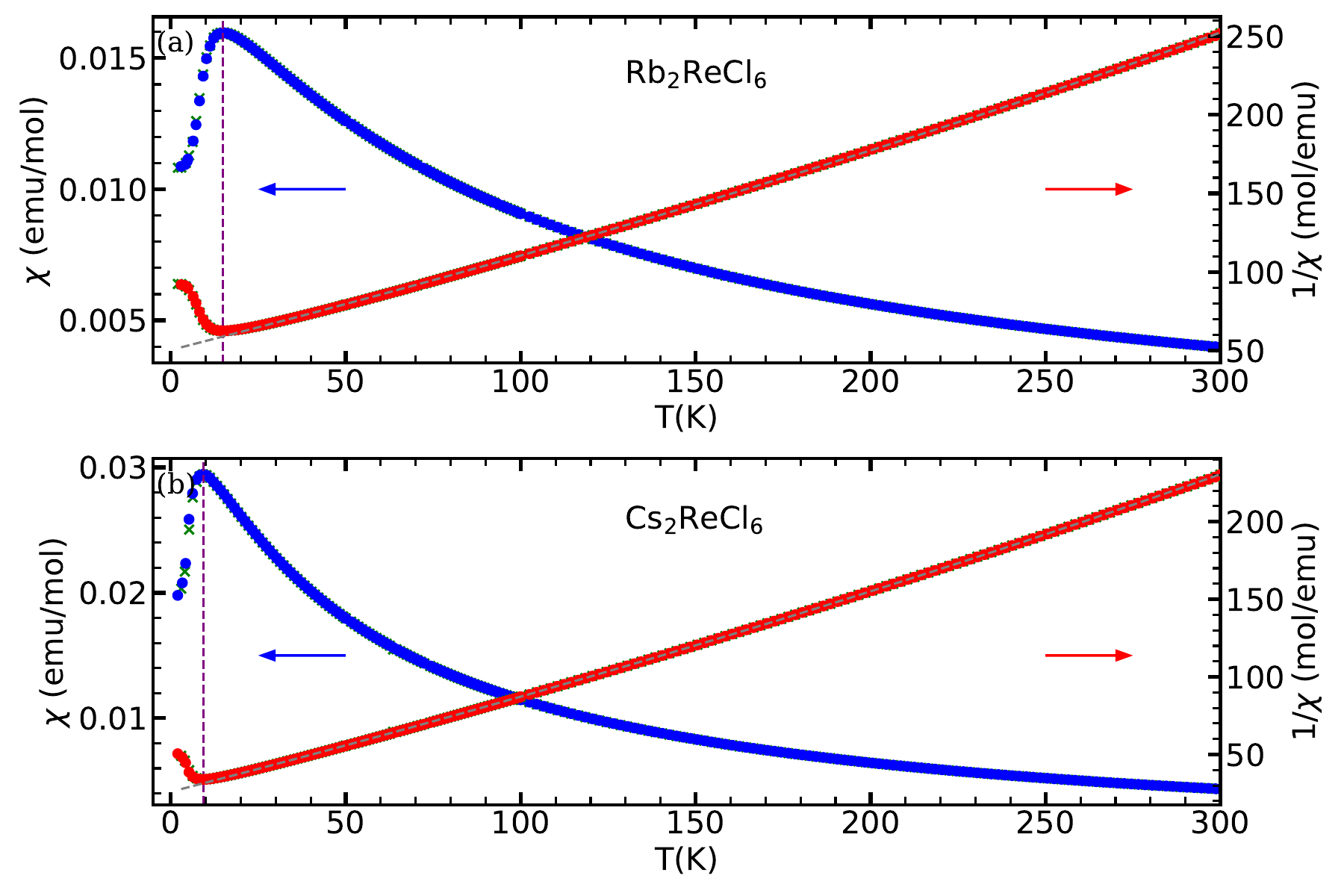}
	\caption{\label{squid} Temperature dependent field-cooled susceptibility (blue) and its inverse (red) determined by SQUID measurements at 0.1\,T for  Rb$_2$ReCl$_6$ (a) and Cs$_2$ReCl$_6$ (b) upon heating. The susceptibility data were fitted by a Curie-Weiss law as described in the text. The green $\times$ symbols plotted below the blue ones denote zero-field-cooled measurements but there is no difference with the results obtained upon field cooling.}
\end{figure}

SQUID experiments determine the  magnetic properties of Rb$_2$ReCl$_6$ and Cs$_2$ReCl$_6$; see Fig.~\ref{squid}. The drop of the susceptibility at low temperature indicates the occurrence of
antiferromagnetic order at 	$T_\mathrm{N}$=15.0(6)\,K and 9.4(6)\,K, respectively. The lower antiferromagnetic transition
in the Cs compound can be attributed to the larger lattice constant inducing smaller magnetic coupling.
However, both values are comparable to the  $T_\mathrm{N}$=12\,K in \krc , which indicates that the 
structural distortions of the Re compound do not lift frustration with an essential impact on the magnetic ordering. The antiferromagnetic order must essentially be stabilized by next-nearest and even farther-neighbor interactions that are not frustrated in the face-centered cubic structure. 

The temperature dependence of the magnetic molar susceptibility was fitted by a Curie-Weiss law:

\begin{equation}\label{eq:CW}
	\chi(T)= N_{\rm A} \frac{(p_{\mathrm{eff}}\mu_{\mathrm B})^2}{3k_{\rm B}(T-\theta_W)}+\chi_0 \, ,
\end{equation}

where, $N_{\rm A}$  is the Avogadro constant, $k_{\rm B}$ the Boltzmann constant, $p_{\mathrm{eff}}$ the effective magnetic moment in Bohr magnetons, $\theta_W$ the Weiss temperature and $\chi_0$ a temperature independent background. The effective moments are close to the free-fitted value obtained in \krc , 3.98\,$\mu_{\mathrm B}$ \cite{Bertin24a}, and all three experimental moments agree within a few percent with the value expected for a pure spin 3/2 moment, 3.87\,$\mu_{\mathrm B}$, which suggests only a small impact of the spin-orbit coupling. 
On a mean-field level the Weiss temperature is given by $\theta_W=\sum_{i}{\frac{1}{3}z_iJ_iS(S+1)}$, where $J_i$ is the Heisenberg interaction per bond (in Kelvin) and $z_i$ the number of neighbors.  
Comparing the Weiss temperatures given in Tab. II with that of \krc , -112\,K \cite{Bertin24a}, there is no indication 
for particularly enhanced antiferromagnetic interaction in the structurally distorted \krc ~ compound, 
but all three Weiss temperatures seem to follow the increase in lattice volume. 
The rather large values of the Weiss temperature can be understood with only moderate nearest-neighbor interaction due to the large spin moment combined with $z_1$=12.
Inspecting the 
ratio between N\'eel and Weiss temperatures, there is no indication for enhanced frustration 
in the cubic materials.

Figure~\ref{xrayStoe}(c) compares the x-ray diffraction patterns obtained for Rb$_2$ReCl$_6$ at $T_\mathrm{N}$=15\,K and below
in the aim to search for some splitting induced by the antiferromagnetic order that can reduce the symmetry similar to the triclinic
distortion emerging in the antiferromagnetic phase of \krc ; see Fig.~5. Due to the negligible thermal expansion 
in this temperature range we can simply subtract these two patterns [red curve in Fig.~\ref{xrayStoe}(c)] obtaining essentially a zero line. There are only tiny indications for peak broadening in the antiferromagnetic state.

In summary the studies of Rb$_2$ReCl$_6$ and Cs$_2$ReCl$_6$ reveal quite similar magnetic properties
compared to \krc , indicating that the rotational distortions do not severely impact the magnetic 
coupling, which seems to follow the lattice enlargement implied by the larger alkaline ions.
Furthermore, the effective moments agree with the expectation for a pure spin moment in the three 
materials suggesting only moderate impact of spin-orbit coupling.

\section{Conclusion}

The two antifluorite materials \krc \ and \ksc \ exhibit the same sequence of structural phase transitions associated with rotation and tilting of the octahedron around different axes parallel to either a $Me$-Cl bond or a Cl-Cl octahedron edge with either the same phase at all sites or with staggering of  opposite rotation angles. Although these structural deformations resemble those observed in oxide perovskites in the GdFeO$_3$ structure type, the complexity in the antifluorites is larger due to the double-perovskite
character with an empty $B^\prime$ site. This complexity impedes the search for an
electronically or Jahn-Teller driven deformation of the $Me$Cl$_6$ octahedron. Crystal structure investigations were performed by high-resolution PND and by SXD and agree very well with each other. The occurrence of the same sequence of structural phases in the two materials with a partially filled and a completely filled $d$ shell documents that these
transitions are not driven by electronic effects in the $d$ shell but arise from the
common bond-length mismatch that is well established for perovskites.
While both materials exhibit some internal deformations of the $Me$Cl$_6$ octahedron
at intermediate temperatures, we find a rather ideal octahedron at low temperatures in the second monoclinic phase in both compounds. A sizable electronic or Jahn-Teller effect can thus be excluded, and 
the intermediate distortions seem to stem from internal strain in these rotated phases.

The low-temperature crystal structure in both compounds exhibits well-defined long-range ordered
rotation and tilt deformations, but evidence for some local disorder can be detected for \ksc \ in the high-temperature cubic phase just above the upper structural transition. The magnetoelastic coupling in \krc \ can be quantified by high-resolution PND studies, which reveal that the degeneracy of magnetic interaction is lifted by the triclinic distortion of the lattice. 

The main conclusion about the absence of a Jahn-Teller distortion in \krc ~ is further supported by the analysis of the two materials with isoelectronic ReCl$_6$ groups
Rb$_2$ReCl$_6$ and Cs$_2$ReCl$_6$, in which larger alkaline ions suppress the rotational
phase transitions. Indeed we do not find any evidence for such transitions, and the
cubic lattice symmetry persisting to low temperatures renders a Jahn-Teller effect very unlikely also for
these materials with identical electronic configuration. The magnetic properties including
the occurrence of antiferromagnetic order are very similar to those in \krc ~ indicating
an important impact of more distant and non-frustrated interaction parameters.

~

\begin{acknowledgments}
	We acknowledge support by the DFG (German Research Foundation) via Project No. 277146847-CRC 1238 (Subprojects A02 and B04). This work is
	partially based on experiments performed at the Swiss spallation
	neutron source SINQ, Paul Scherrer Institute, Villigen,
	Switzerland. We thank the ILL, Grenoble, France, for access to the D8-Venture instrument. The supporting data and codes for this article are available from Zenodo.
\end{acknowledgments}


\begin{thebibliography}{66}%
	\makeatletter
	\providecommand \@ifxundefined [1]{%
		\@ifx{#1\undefined}
	}%
	\providecommand \@ifnum [1]{%
		\ifnum #1\expandafter \@firstoftwo
		\else \expandafter \@secondoftwo
		\fi
	}%
	\providecommand \@ifx [1]{%
		\ifx #1\expandafter \@firstoftwo
		\else \expandafter \@secondoftwo
		\fi
	}%
	\providecommand \natexlab [1]{#1}%
	\providecommand \enquote  [1]{``#1''}%
	\providecommand \bibnamefont  [1]{#1}%
	\providecommand \bibfnamefont [1]{#1}%
	\providecommand \citenamefont [1]{#1}%
	\providecommand \href@noop [0]{\@secondoftwo}%
	\providecommand \href [0]{\begingroup \@sanitize@url \@href}%
	\providecommand \@href[1]{\@@startlink{#1}\@@href}%
	\providecommand \@@href[1]{\endgroup#1\@@endlink}%
	\providecommand \@sanitize@url [0]{\catcode `\\12\catcode `\$12\catcode
		`\&12\catcode `\#12\catcode `\^12\catcode `\_12\catcode `\%12\relax}%
	\providecommand \@@startlink[1]{}%
	\providecommand \@@endlink[0]{}%
	\providecommand \url  [0]{\begingroup\@sanitize@url \@url }%
	\providecommand \@url [1]{\endgroup\@href {#1}{\urlprefix }}%
	\providecommand \urlprefix  [0]{URL }%
	\providecommand \Eprint [0]{\href }%
	\providecommand \doibase [0]{https://doi.org/}%
	\providecommand \selectlanguage [0]{\@gobble}%
	\providecommand \bibinfo  [0]{\@secondoftwo}%
	\providecommand \bibfield  [0]{\@secondoftwo}%
	\providecommand \translation [1]{[#1]}%
	\providecommand \BibitemOpen [0]{}%
	\providecommand \bibitemStop [0]{}%
	\providecommand \bibitemNoStop [0]{.\EOS\space}%
	\providecommand \EOS [0]{\spacefactor3000\relax}%
	\providecommand \BibitemShut  [1]{\csname bibitem#1\endcsname}%
	\let\auto@bib@innerbib\@empty
	%</preamble>
	\bibitem [{\citenamefont {Raghu}\ \emph {et~al.}(2008)\citenamefont {Raghu},
		\citenamefont {Qi}, \citenamefont {Honerkamp},\ and\ \citenamefont
		{Zhang}}]{Raghu2008}%
	\BibitemOpen
	\bibfield  {author} {\bibinfo {author} {\bibfnamefont {S.}~\bibnamefont
			{Raghu}}, \bibinfo {author} {\bibfnamefont {X.-L.}\ \bibnamefont {Qi}},
		\bibinfo {author} {\bibfnamefont {C.}~\bibnamefont {Honerkamp}},\ and\
		\bibinfo {author} {\bibfnamefont {S.-C.}\ \bibnamefont {Zhang}},\ }\bibfield
	{title} {\bibinfo {title} {Topological mott insulators},\ }\href@noop {}
	{\bibfield  {journal} {\bibinfo  {journal} {Phys. Rev. Lett.}\
		}\textbf {\bibinfo {volume} {100}},\ \bibinfo {pages} {156401} (\bibinfo
		{year} {2008})}\BibitemShut {NoStop}%
	\bibitem [{\citenamefont {Zhang}\ \emph {et~al.}(2012)\citenamefont {Zhang},
		\citenamefont {Zhang}, \citenamefont {Wang}, \citenamefont {Felser},\ and\
		\citenamefont {Zhang}}]{Zhang2012}%
	\BibitemOpen
	\bibfield  {author} {\bibinfo {author} {\bibfnamefont {X.}~\bibnamefont
			{Zhang}}, \bibinfo {author} {\bibfnamefont {H.}~\bibnamefont {Zhang}},
		\bibinfo {author} {\bibfnamefont {J.}~\bibnamefont {Wang}}, \bibinfo {author}
		{\bibfnamefont {C.}~\bibnamefont {Felser}},\ and\ \bibinfo {author}
		{\bibfnamefont {S.-C.}\ \bibnamefont {Zhang}},\ }\bibfield  {title} {\bibinfo
		{title} {Actinide topological insulator materials with strong interaction},\
	}\href@noop {} {\bibfield  {journal} {\bibinfo  {journal} {Science}\ }\textbf
		{\bibinfo {volume} {335}},\ \bibinfo {pages} {1464} (\bibinfo {year}
		{2012})}\BibitemShut {NoStop}%
	\bibitem [{\citenamefont {Maciejko}\ and\ \citenamefont
		{Fiete}(2015)}]{Maciejko2015}%
	\BibitemOpen
	\bibfield  {author} {\bibinfo {author} {\bibfnamefont {J.}~\bibnamefont
			{Maciejko}}\ and\ \bibinfo {author} {\bibfnamefont {G.~A.}\ \bibnamefont
			{Fiete}},\ }\bibfield  {title} {\bibinfo {title} {Fractionalized topological
			insulators},\ }\href@noop {} {\bibfield  {journal} {\bibinfo  {journal}
			{Nature Physics}\ }\textbf {\bibinfo {volume} {11}},\ \bibinfo {pages} {385}
		(\bibinfo {year} {2015})}\BibitemShut {NoStop}%
	\bibitem [{\citenamefont {Trebst}\ and\ \citenamefont
		{Hickey}(2022)}]{Trebst2022}%
	\BibitemOpen
	\bibfield  {author} {\bibinfo {author} {\bibfnamefont {S.}~\bibnamefont
			{Trebst}}\ and\ \bibinfo {author} {\bibfnamefont {C.}~\bibnamefont
			{Hickey}},\ }\bibfield  {title} {\bibinfo {title} {{Kitaev materials}},\
	}\href {https://doi.org/https://doi.org/10.1016/j.physrep.2021.11.003}
	{\bibfield  {journal} {\bibinfo  {journal} {Physics Reports}\ }\textbf
		{\bibinfo {volume} {950}},\ \bibinfo {pages} {1} (\bibinfo {year}
		{2022})}\BibitemShut {NoStop}%
	\bibitem [{\citenamefont {Takagi}\ \emph {et~al.}(2019)\citenamefont {Takagi},
		\citenamefont {Takayama}, \citenamefont {Jackeli}, \citenamefont
		{Khaliullin},\ and\ \citenamefont {Nagler}}]{takagi2019}%
	\BibitemOpen
	\bibfield  {author} {\bibinfo {author} {\bibfnamefont {H.}~\bibnamefont
			{Takagi}}, \bibinfo {author} {\bibfnamefont {T.}~\bibnamefont {Takayama}},
		\bibinfo {author} {\bibfnamefont {G.}~\bibnamefont {Jackeli}}, \bibinfo
		{author} {\bibfnamefont {G.}~\bibnamefont {Khaliullin}},\ and\ \bibinfo
		{author} {\bibfnamefont {S.~E.}\ \bibnamefont {Nagler}},\ }\bibfield  {title}
	{\bibinfo {title} {{Concept and realization of Kitaev quantum spin
				liquids}},\ }\href@noop {} {\bibfield  {journal} {\bibinfo  {journal} {Nature
				Reviews Physics}\ }\textbf {\bibinfo {volume} {1}},\ \bibinfo {pages} {264}
		(\bibinfo {year} {2019})}\BibitemShut {NoStop}%
	\bibitem [{\citenamefont {Kitaev}(2006)}]{Kitaev06}%
	\BibitemOpen
	\bibfield  {author} {\bibinfo {author} {\bibfnamefont {A.}~\bibnamefont
			{Kitaev}},\ }\bibfield  {title} {\bibinfo {title} {{Anyons in an exactly
				solved model and beyond}},\ }\href
	{https://doi.org/https://doi.org/10.1016/j.aop.2005.10.005} {\bibfield
		{journal} {\bibinfo  {journal} {Annals of Physics}\ }\textbf {\bibinfo
			{volume} {321}},\ \bibinfo {pages} {2} (\bibinfo {year} {2006})},\ \bibinfo
	{note} {january Special Issue}\BibitemShut {NoStop}%
	\bibitem [{\citenamefont {Khomskii}(2014)}]{Khomskii14}%
	\BibitemOpen
	\bibfield  {author} {\bibinfo {author} {\bibfnamefont {D.~I.}\ \bibnamefont
			{Khomskii}},\ }\href@noop {} {\emph {\bibinfo {title} {Transition Metal
				Compounds}}}\ (\bibinfo  {publisher} {Cambridge University Press},\ \bibinfo
	{year} {2014})\BibitemShut {NoStop}%
	\bibitem [{\citenamefont {Streltsov}\ and\ \citenamefont
		{Khomskii}(2020{\natexlab{a}})}]{Streltsov20}%
	\BibitemOpen
	\bibfield  {author} {\bibinfo {author} {\bibfnamefont {S.~V.}\ \bibnamefont
			{Streltsov}}\ and\ \bibinfo {author} {\bibfnamefont {D.~I.}\ \bibnamefont
			{Khomskii}},\ }\bibfield  {title} {\bibinfo {title} {{Jahn-Teller Effect and
				Spin-Orbit Coupling: Friends or Foes?}},\ }\href
	{https://doi.org/10.1103/PhysRevX.10.031043} {\bibfield  {journal} {\bibinfo
			{journal} {Phys. Rev. X}\ }\textbf {\bibinfo {volume} {10}},\ \bibinfo
		{pages} {031043} (\bibinfo {year} {2020}{\natexlab{a}})}\BibitemShut
	{NoStop}%
	\bibitem [{\citenamefont {Goodenough}(1963)}]{Goodenough63}%
	\BibitemOpen
	\bibfield  {author} {\bibinfo {author} {\bibfnamefont {J.}~\bibnamefont
			{Goodenough}},\ }\href {https://books.google.de/books?id=ljtRAAAAMAAJ} {\emph
		{\bibinfo {title} {Magnetism and the Chemical Bond}}},\ Inorganic Chemistry
	Section / Interscience monographs on chemistry\ (\bibinfo  {publisher}
	{Interscience Publishers},\ \bibinfo {year} {1963})\BibitemShut {NoStop}%
	\bibitem [{\citenamefont {Ishikawa}\ \emph {et~al.}(2019)\citenamefont
		{Ishikawa}, \citenamefont {Takayama}, \citenamefont {Kremer}, \citenamefont
		{Nuss}, \citenamefont {Dinnebier}, \citenamefont {Kitagawa}, \citenamefont
		{Ishii},\ and\ \citenamefont {Takagi}}]{Ishikawa19}%
	\BibitemOpen
	\bibfield  {author} {\bibinfo {author} {\bibfnamefont {H.}~\bibnamefont
			{Ishikawa}}, \bibinfo {author} {\bibfnamefont {T.}~\bibnamefont {Takayama}},
		\bibinfo {author} {\bibfnamefont {R.~K.}\ \bibnamefont {Kremer}}, \bibinfo
		{author} {\bibfnamefont {J.}~\bibnamefont {Nuss}}, \bibinfo {author}
		{\bibfnamefont {R.}~\bibnamefont {Dinnebier}}, \bibinfo {author}
		{\bibfnamefont {K.}~\bibnamefont {Kitagawa}}, \bibinfo {author}
		{\bibfnamefont {K.}~\bibnamefont {Ishii}},\ and\ \bibinfo {author}
		{\bibfnamefont {H.}~\bibnamefont {Takagi}},\ }\bibfield  {title} {\bibinfo
		{title} {{Ordering of hidden multipoles in spin-orbit entangled $5{d}^{1}$ Ta
				chlorides}},\ }\href {https://doi.org/10.1103/PhysRevB.100.045142} {\bibfield
		{journal} {\bibinfo  {journal} {Phys. Rev. B}\ }\textbf {\bibinfo {volume}
			{100}},\ \bibinfo {pages} {045142} (\bibinfo {year} {2019})}\BibitemShut
	{NoStop}%
	\bibitem [{\citenamefont {Khomskii}\ and\ \citenamefont
		{Streltsov}(2021)}]{Khomskii21}%
	\BibitemOpen
	\bibfield  {author} {\bibinfo {author} {\bibfnamefont {D.~I.}\ \bibnamefont
			{Khomskii}}\ and\ \bibinfo {author} {\bibfnamefont {S.~V.}\ \bibnamefont
			{Streltsov}},\ }\bibfield  {title} {\bibinfo {title} {{Orbital Effects in
				Solids: Basics, Recent Progress, and Opportunities}},\ }\href
	{https://doi.org/10.1021/acs.chemrev.0c00579} {\bibfield  {journal} {\bibinfo
			{journal} {Chemical Reviews}\ }\textbf {\bibinfo {volume} {121}},\ \bibinfo
		{pages} {2992} (\bibinfo {year} {2021})},\ \bibinfo {note} {pMID:
		33314912}\BibitemShut {NoStop}%
	\bibitem [{\citenamefont {Khomskii}\ and\ \citenamefont {van~den
			Brink}(2000)}]{Khomskii00}%
	\BibitemOpen
	\bibfield  {author} {\bibinfo {author} {\bibfnamefont {D.}~\bibnamefont
			{Khomskii}}\ and\ \bibinfo {author} {\bibfnamefont {J.}~\bibnamefont {van~den
				Brink}},\ }\bibfield  {title} {\bibinfo {title} {{Anharmonic Effects on
				Charge and Orbital Order}},\ }\href
	{https://doi.org/10.1103/PhysRevLett.85.3329} {\bibfield  {journal} {\bibinfo
			{journal} {Phys. Rev. Lett.}\ }\textbf {\bibinfo {volume} {85}},\ \bibinfo
		{pages} {3329} (\bibinfo {year} {2000})}\BibitemShut {NoStop}%
	\bibitem [{\citenamefont {Weng}\ and\ \citenamefont {Dong}(2021)}]{Weng21}%
	\BibitemOpen
	\bibfield  {author} {\bibinfo {author} {\bibfnamefont {Y.}~\bibnamefont
			{Weng}}\ and\ \bibinfo {author} {\bibfnamefont {S.}~\bibnamefont {Dong}},\
	}\bibfield  {title} {\bibinfo {title} {{Manipulation of
				${J}_{\mathrm{eff}}\phantom{\rule{4pt}{0ex}}=\phantom{\rule{4pt}{0ex}}\frac{3}{2}$
				states by tuning the tetragonal distortion}},\ }\href
	{https://doi.org/10.1103/PhysRevB.104.165150} {\bibfield  {journal} {\bibinfo
			{journal} {Phys. Rev. B}\ }\textbf {\bibinfo {volume} {104}},\ \bibinfo
		{pages} {165150} (\bibinfo {year} {2021})}\BibitemShut {NoStop}%
	\bibitem [{\citenamefont {Yuan}\ \emph {et~al.}(2017)\citenamefont {Yuan},
		\citenamefont {Clancy}, \citenamefont {Cook}, \citenamefont {Thompson},
		\citenamefont {Greedan}, \citenamefont {Cao}, \citenamefont {Jeon},
		\citenamefont {Noh}, \citenamefont {Upton}, \citenamefont {Casa},
		\citenamefont {Gog}, \citenamefont {Paramekanti},\ and\ \citenamefont
		{Kim}}]{Yuan17}%
	\BibitemOpen
	\bibfield  {author} {\bibinfo {author} {\bibfnamefont {B.}~\bibnamefont
			{Yuan}}, \bibinfo {author} {\bibfnamefont {J.~P.}\ \bibnamefont {Clancy}},
		\bibinfo {author} {\bibfnamefont {A.~M.}\ \bibnamefont {Cook}}, \bibinfo
		{author} {\bibfnamefont {C.~M.}\ \bibnamefont {Thompson}}, \bibinfo {author}
		{\bibfnamefont {J.}~\bibnamefont {Greedan}}, \bibinfo {author} {\bibfnamefont
			{G.}~\bibnamefont {Cao}}, \bibinfo {author} {\bibfnamefont {B.~C.}\
			\bibnamefont {Jeon}}, \bibinfo {author} {\bibfnamefont {T.~W.}\ \bibnamefont
			{Noh}}, \bibinfo {author} {\bibfnamefont {M.~H.}\ \bibnamefont {Upton}},
		\bibinfo {author} {\bibfnamefont {D.}~\bibnamefont {Casa}}, \bibinfo {author}
		{\bibfnamefont {T.}~\bibnamefont {Gog}}, \bibinfo {author} {\bibfnamefont
			{A.}~\bibnamefont {Paramekanti}},\ and\ \bibinfo {author} {\bibfnamefont
			{Y.-J.}\ \bibnamefont {Kim}},\ }\bibfield  {title} {\bibinfo {title}
		{{Determination of Hund's coupling in $5d$ oxides using resonant inelastic
				x-ray scattering}},\ }\href {https://doi.org/10.1103/PhysRevB.95.235114}
	{\bibfield  {journal} {\bibinfo  {journal} {Phys. Rev. B}\ }\textbf {\bibinfo
			{volume} {95}},\ \bibinfo {pages} {235114} (\bibinfo {year}
		{2017})}\BibitemShut {NoStop}%
	\bibitem [{\citenamefont {Soh}\ \emph {et~al.}(2024)\citenamefont {Soh},
		\citenamefont {Merkel}, \citenamefont {Pourovskii}, \citenamefont
		{{\v{Z}}ivkovi{\'c}}, \citenamefont {Malanyuk}, \citenamefont
		{P{\'a}sztorov{\'a}}, \citenamefont {Francoual}, \citenamefont {Hirai},
		\citenamefont {Urru}, \citenamefont {Tolj} \emph {et~al.}}]{Soh24}%
	\BibitemOpen
	\bibfield  {author} {\bibinfo {author} {\bibfnamefont {J.-R.}\ \bibnamefont
			{Soh}}, \bibinfo {author} {\bibfnamefont {M.~E.}\ \bibnamefont {Merkel}},
		\bibinfo {author} {\bibfnamefont {L.~V.}\ \bibnamefont {Pourovskii}},
		\bibinfo {author} {\bibfnamefont {I.}~\bibnamefont {{\v{Z}}ivkovi{\'c}}},
		\bibinfo {author} {\bibfnamefont {O.}~\bibnamefont {Malanyuk}}, \bibinfo
		{author} {\bibfnamefont {J.}~\bibnamefont {P{\'a}sztorov{\'a}}}, \bibinfo
		{author} {\bibfnamefont {S.}~\bibnamefont {Francoual}}, \bibinfo {author}
		{\bibfnamefont {D.}~\bibnamefont {Hirai}}, \bibinfo {author} {\bibfnamefont
			{A.}~\bibnamefont {Urru}}, \bibinfo {author} {\bibfnamefont {D.}~\bibnamefont
			{Tolj}}, \emph {et~al.},\ }\bibfield  {title} {\bibinfo {title}
		{{Spectroscopic signatures and origin of hidden order in Ba$_2$MgRe$_6$}},\
	}\href@noop {} {\bibfield  {journal} {\bibinfo  {journal} {Nature
				Communications}\ }\textbf {\bibinfo {volume} {15}},\ \bibinfo {pages} {1}
		(\bibinfo {year} {2024})}\BibitemShut {NoStop}%
	\bibitem [{\citenamefont {Streltsov}\ and\ \citenamefont
		{Khomskii}(2020{\natexlab{b}})}]{Streltsov2020}%
	\BibitemOpen
	\bibfield  {author} {\bibinfo {author} {\bibfnamefont {S.~V.}\ \bibnamefont
			{Streltsov}}\ and\ \bibinfo {author} {\bibfnamefont {D.~I.}\ \bibnamefont
			{Khomskii}},\ }\bibfield  {title} {\bibinfo {title} {{Jahn-Teller Effect and
				Spin-Orbit Coupling: Friends or Foes?}},\ }\href
	{https://doi.org/10.1103/PhysRevX.10.031043} {\bibfield  {journal} {\bibinfo
			{journal} {Phys. Rev. X}\ }\textbf {\bibinfo {volume} {10}},\ \bibinfo
		{pages} {031043} (\bibinfo {year} {2020}{\natexlab{b}})}\BibitemShut
	{NoStop}%
	\bibitem [{\citenamefont {{Warzanowski, P. and Magnaterra, M. and Schlicht, G.
				and Faure, Q. and Sahle, Ch. J. and Becker, P. and Bohat\'y, L. and Sala, M.
				Moretti and Monaco, G. and Hermanns, M. and van Loosdrecht, P. H. M. and
				Gr\"uninger, M.}}(2024)}]{Warzanowski24}%
	\BibitemOpen
	\bibfield  {author} {\bibinfo {author} {\bibnamefont {{Warzanowski, P. and
					Magnaterra, M. and Schlicht, G. and Faure, Q. and Sahle, Ch. J. and Becker,
					P. and Bohat\'y, L. and Sala, M. Moretti and Monaco, G. and Hermanns, M. and
					van Loosdrecht, P. H. M. and Gr\"uninger, M.}}},\ }\bibfield  {title}
	{\bibinfo {title} {{Spin-orbit coupling in a half-filled ${t}_{2g}$ shell:
				The case of $5{d}^{3}$ ${\mathrm{K}}_{2}{\mathrm{ReCl}}_{6}$}},\ }\href
	{https://doi.org/10.1103/PhysRevB.109.155149} {\bibfield  {journal} {\bibinfo
			{journal} {Phys. Rev. B}\ }\textbf {\bibinfo {volume} {109}},\ \bibinfo
		{pages} {155149} (\bibinfo {year} {2024})}\BibitemShut {NoStop}%
	\bibitem [{\citenamefont {Du}\ \emph {et~al.}(2025)\citenamefont {Du},
		\citenamefont {Hao}, \citenamefont {Hao}, \citenamefont {Jia}, \citenamefont
		{Sun},\ and\ \citenamefont {Xu}}]{Du25}%
	\BibitemOpen
	\bibfield  {author} {\bibinfo {author} {\bibfnamefont {Y.}~\bibnamefont
			{Du}}, \bibinfo {author} {\bibfnamefont {Y.}~\bibnamefont {Hao}}, \bibinfo
		{author} {\bibfnamefont {X.}~\bibnamefont {Hao}}, \bibinfo {author}
		{\bibfnamefont {Y.}~\bibnamefont {Jia}}, \bibinfo {author} {\bibfnamefont
			{K.}~\bibnamefont {Sun}},\ and\ \bibinfo {author} {\bibfnamefont
			{Y.}~\bibnamefont {Xu}},\ }\bibfield  {title} {\bibinfo {title} {{Magnetic
				properties and spin–orbit coupling-driven Jahn–Teller distortions in
				K$_2$ReX$_6$ (X = Cl{,} Br and I) with a half-filled 5$d$-$t^3_{2g}$
				shell}},\ }\href {https://doi.org/10.1039/D5CP02013A} {\bibfield  {journal}
		{\bibinfo  {journal} {Phys. Chem. Chem. Phys.}\ }\textbf {\bibinfo {volume}
			{27}},\ \bibinfo {pages} {20699} (\bibinfo {year} {2025})}\BibitemShut
	{NoStop}%
	\bibitem [{\citenamefont {Momma}\ and\ \citenamefont
		{Izumi}(2011)}]{Momma2011}%
	\BibitemOpen
	\bibfield  {author} {\bibinfo {author} {\bibfnamefont {K.}~\bibnamefont
			{Momma}}\ and\ \bibinfo {author} {\bibfnamefont {F.}~\bibnamefont {Izumi}},\
	}\bibfield  {title} {\bibinfo {title} {{{\it VESTA3} for three-dimensional
				visualization of crystal, volumetric and morphology data}},\ }\href
	{https://doi.org/10.1107/S0021889811038970} {\bibfield  {journal} {\bibinfo
			{journal} {Journal of Applied Crystallography}\ }\textbf {\bibinfo {volume}
			{44}},\ \bibinfo {pages} {1272} (\bibinfo {year} {2011})}\BibitemShut
	{NoStop}%
	\bibitem [{\citenamefont {Busey}\ and\ \citenamefont
		{Sonder}(1962)}]{Busey1962}%
	\BibitemOpen
	\bibfield  {author} {\bibinfo {author} {\bibfnamefont {R.~H.}\ \bibnamefont
			{Busey}}\ and\ \bibinfo {author} {\bibfnamefont {E.}~\bibnamefont {Sonder}},\
	}\bibfield  {title} {\bibinfo {title} {Magnetic {S}usceptibility of
			{P}otassium {H}exachlororhenate ({IV}) and {P}otassium {H}exabromorhenate
			({IV}) from 5$^\circ$ to 300$^\circ${K}},\ }\href
	{https://doi.org/10.1063/1.1732325} {\bibfield  {journal} {\bibinfo
			{journal} {The Journal of Chemical Physics}\ }\textbf {\bibinfo {volume}
			{36}},\ \bibinfo {pages} {93} (\bibinfo {year} {1962})}\BibitemShut {NoStop}%
	\bibitem [{\citenamefont {Busey}\ \emph {et~al.}(1962)\citenamefont {Busey},
		\citenamefont {Dearman},\ and\ \citenamefont {Bevan}}]{Busey1962a}%
	\BibitemOpen
	\bibfield  {author} {\bibinfo {author} {\bibfnamefont {R.~H.}\ \bibnamefont
			{Busey}}, \bibinfo {author} {\bibfnamefont {H.~H.}\ \bibnamefont {Dearman}},\
		and\ \bibinfo {author} {\bibfnamefont {R.~B.}\ \bibnamefont {Bevan}},\
	}\bibfield  {title} {\bibinfo {title} {The heat capacity of potassium
			hexachlororhenate({IV}) from 7 to 320 {K}. {A}nomalies near 12, 76, 103, and
			111 {K}. {E}ntropy and free energy functions. {S}olubility and heat of
			solution of {K}$_2${R}e{C}l$_6$. {E}ntropy of the hexachlororhenate ion},\
	}\href {https://doi.org/10.1021/j100807a017} {\bibfield  {journal} {\bibinfo
			{journal} {The Journal of Physical Chemistry}\ }\textbf {\bibinfo {volume}
			{66}},\ \bibinfo {pages} {82} (\bibinfo {year} {1962})}\BibitemShut {NoStop}%
	\bibitem [{\citenamefont {O'Leary}\ and\ \citenamefont
		{Wheeler}(1970)}]{OLeary1970}%
	\BibitemOpen
	\bibfield  {author} {\bibinfo {author} {\bibfnamefont {G.~P.}\ \bibnamefont
			{O'Leary}}\ and\ \bibinfo {author} {\bibfnamefont {R.~G.}\ \bibnamefont
			{Wheeler}},\ }\bibfield  {title} {\bibinfo {title} {{P}hase {T}ransitions and
			{S}oft {L}ibrational {M}odes in {C}ubic {C}rystals},\ }\href
	{https://doi.org/10.1103/PhysRevB.1.4409} {\bibfield  {journal} {\bibinfo
			{journal} {Phys. Rev. B}\ }\textbf {\bibinfo {volume} {1}},\ \bibinfo {pages}
		{4409} (\bibinfo {year} {1970})}\BibitemShut {NoStop}%
	\bibitem [{\citenamefont {Armstrong}(1980{\natexlab{a}})}]{Armstrong1980}%
	\BibitemOpen
	\bibfield  {author} {\bibinfo {author} {\bibfnamefont {R.~L.}\ \bibnamefont
			{Armstrong}},\ }\bibfield  {title} {\bibinfo {title} {Structural properties
			and lattice dynamics of 5d transition metal antifluorite crystals},\ }\href
	{https://doi.org/https://doi.org/10.1016/0370-1573(80)90146-5} {\bibfield
		{journal} {\bibinfo  {journal} {Physics Reports}\ }\textbf {\bibinfo {volume}
			{57}},\ \bibinfo {pages} {343} (\bibinfo {year}
		{1980}{\natexlab{a}})}\BibitemShut {NoStop}%
	\bibitem [{\citenamefont {Willemsen}\ \emph
		{et~al.}(1977{\natexlab{a}})\citenamefont {Willemsen}, \citenamefont
		{Martin}, \citenamefont {Meincke},\ and\ \citenamefont
		{Armstrong}}]{Willemsen1977}%
	\BibitemOpen
	\bibfield  {author} {\bibinfo {author} {\bibfnamefont {H.~W.}\ \bibnamefont
			{Willemsen}}, \bibinfo {author} {\bibfnamefont {C.~A.}\ \bibnamefont
			{Martin}}, \bibinfo {author} {\bibfnamefont {P.~P.~M.}\ \bibnamefont
			{Meincke}},\ and\ \bibinfo {author} {\bibfnamefont {R.~L.}\ \bibnamefont
			{Armstrong}},\ }\bibfield  {title} {\bibinfo {title} {{Thermal-expansion
				study of the displacive phase transitions in {K}$_2${R}e{C}l$_6$ and
				{K}$_2${O}s{C}l$_6$}},\ }\href {https://doi.org/10.1103/PhysRevB.16.2283}
	{\bibfield  {journal} {\bibinfo  {journal} {Phys. Rev. B}\ }\textbf {\bibinfo
			{volume} {16}},\ \bibinfo {pages} {2283} (\bibinfo {year}
		{1977}{\natexlab{a}})}\BibitemShut {NoStop}%
	\bibitem [{\citenamefont {Willemsen}\ \emph
		{et~al.}(1977{\natexlab{b}})\citenamefont {Willemsen}, \citenamefont
		{Armstrong},\ and\ \citenamefont {Meincke}}]{Willemsen1977a}%
	\BibitemOpen
	\bibfield  {author} {\bibinfo {author} {\bibfnamefont {H.~W.}\ \bibnamefont
			{Willemsen}}, \bibinfo {author} {\bibfnamefont {R.~L.}\ \bibnamefont
			{Armstrong}},\ and\ \bibinfo {author} {\bibfnamefont {P.~P.}\ \bibnamefont
			{Meincke}},\ }\bibfield  {title} {\bibinfo {title} {{Thermal expansion
				measurements near the antiferromagnetic phase transitions in K$_2$ReCl$_6$
				and K$_2$IrCl$_6$}},\ }\href@noop {} {\bibfield  {journal} {\bibinfo
			{journal} {Journal of Low Temperature Physics}\ }\textbf {\bibinfo {volume}
			{26}},\ \bibinfo {pages} {299} (\bibinfo {year}
		{1977}{\natexlab{b}})}\BibitemShut {NoStop}%
	\bibitem [{\citenamefont {Smith}\ and\ \citenamefont
		{Bacon}(1966)}]{Smith1966}%
	\BibitemOpen
	\bibfield  {author} {\bibinfo {author} {\bibfnamefont {H.~G.}\ \bibnamefont
			{Smith}}\ and\ \bibinfo {author} {\bibfnamefont {G.~E.}\ \bibnamefont
			{Bacon}},\ }\bibfield  {title} {\bibinfo {title} {{Neutron Diffraction Study
				of Magnetic Ordering in K$_2$ReCl$_6$}},\ }\href
	{https://doi.org/10.1063/1.1708548} {\bibfield  {journal} {\bibinfo
			{journal} {Journal of Applied Physics}\ }\textbf {\bibinfo {volume} {37}},\
		\bibinfo {pages} {979} (\bibinfo {year} {1966})}\BibitemShut {NoStop}%
	\bibitem [{\citenamefont {Minkiewicz}\ \emph {et~al.}(1968)\citenamefont
		{Minkiewicz}, \citenamefont {Shirane}, \citenamefont {Frazer}, \citenamefont
		{Wheeler},\ and\ \citenamefont {Dorain}}]{Minkiewicz1968}%
	\BibitemOpen
	\bibfield  {author} {\bibinfo {author} {\bibfnamefont {V.}~\bibnamefont
			{Minkiewicz}}, \bibinfo {author} {\bibfnamefont {G.}~\bibnamefont {Shirane}},
		\bibinfo {author} {\bibfnamefont {B.}~\bibnamefont {Frazer}}, \bibinfo
		{author} {\bibfnamefont {R.}~\bibnamefont {Wheeler}},\ and\ \bibinfo {author}
		{\bibfnamefont {P.}~\bibnamefont {Dorain}},\ }\bibfield  {title} {\bibinfo
		{title} {{Neutron diffraction study of magnetic ordering in
				{K}$_2${I}r{C}l$_6$, {K}$_2${R}e{B}r$_6$ and {K}$_2${R}e{C}l$_6$}},\ }\href
	{https://doi.org/https://doi.org/10.1016/0022-3697(68)90222-9} {\bibfield
		{journal} {\bibinfo  {journal} {Journal of Physics and Chemistry of Solids}\
		}\textbf {\bibinfo {volume} {29}},\ \bibinfo {pages} {881} (\bibinfo {year}
		{1968})}\BibitemShut {NoStop}%
	\bibitem [{\citenamefont {Bertin}\ \emph
		{et~al.}(2024{\natexlab{a}})\citenamefont {Bertin}, \citenamefont {Dey},
		\citenamefont {Br\"uning}, \citenamefont {Gorkov}, \citenamefont {Jenni},
		\citenamefont {Krause}, \citenamefont {Becker}, \citenamefont {Bohat\'y},
		\citenamefont {Khomskii}, \citenamefont {Pomjakushin}, \citenamefont
		{Keller}, \citenamefont {Braden},\ and\ \citenamefont {Lorenz}}]{Bertin24a}%
	\BibitemOpen
	\bibfield  {author} {\bibinfo {author} {\bibfnamefont {A.}~\bibnamefont
			{Bertin}}, \bibinfo {author} {\bibfnamefont {T.}~\bibnamefont {Dey}},
		\bibinfo {author} {\bibfnamefont {D.}~\bibnamefont {Br\"uning}}, \bibinfo
		{author} {\bibfnamefont {D.}~\bibnamefont {Gorkov}}, \bibinfo {author}
		{\bibfnamefont {K.}~\bibnamefont {Jenni}}, \bibinfo {author} {\bibfnamefont
			{A.}~\bibnamefont {Krause}}, \bibinfo {author} {\bibfnamefont
			{P.}~\bibnamefont {Becker}}, \bibinfo {author} {\bibfnamefont
			{L.}~\bibnamefont {Bohat\'y}}, \bibinfo {author} {\bibfnamefont
			{D.}~\bibnamefont {Khomskii}}, \bibinfo {author} {\bibfnamefont
			{V.}~\bibnamefont {Pomjakushin}}, \bibinfo {author} {\bibfnamefont
			{L.}~\bibnamefont {Keller}}, \bibinfo {author} {\bibfnamefont
			{M.}~\bibnamefont {Braden}},\ and\ \bibinfo {author} {\bibfnamefont
			{T.}~\bibnamefont {Lorenz}},\ }\bibfield  {title} {\bibinfo {title}
		{{Interplay of magnetic order and ferroelasticity in the spin-orbit coupled
				antiferromagnet ${\mathrm{K}}_{2}{\mathrm{ReCl}}_{6}$}},\ }\href
	{https://doi.org/10.1103/PhysRevB.109.094409} {\bibfield  {journal} {\bibinfo
			{journal} {Phys. Rev. B}\ }\textbf {\bibinfo {volume} {109}},\ \bibinfo
		{pages} {094409} (\bibinfo {year} {2024}{\natexlab{a}})}\BibitemShut
	{NoStop}%
	\bibitem [{\citenamefont {Bertin}\ \emph
		{et~al.}(2024{\natexlab{b}})\citenamefont {Bertin}, \citenamefont {Kiefer},
		\citenamefont {Becker}, \citenamefont {Bohatý},\ and\ \citenamefont
		{Braden}}]{Bertin24b}%
	\BibitemOpen
	\bibfield  {author} {\bibinfo {author} {\bibfnamefont {A.}~\bibnamefont
			{Bertin}}, \bibinfo {author} {\bibfnamefont {L.}~\bibnamefont {Kiefer}},
		\bibinfo {author} {\bibfnamefont {P.}~\bibnamefont {Becker}}, \bibinfo
		{author} {\bibfnamefont {L.}~\bibnamefont {Bohatý}},\ and\ \bibinfo {author}
		{\bibfnamefont {M.}~\bibnamefont {Braden}},\ }\bibfield  {title} {\bibinfo
		{title} {{Rotational phase transitions in antifluorite-type osmate and
				iridate compounds}},\ }\href {https://doi.org/10.1088/1361-648X/ad2fef}
	{\bibfield  {journal} {\bibinfo  {journal} {Journal of Physics: Condensed
				Matter}\ }\textbf {\bibinfo {volume} {36}},\ \bibinfo {pages} {245402}
		(\bibinfo {year} {2024}{\natexlab{b}})}\BibitemShut {NoStop}%
	\bibitem [{\citenamefont {Cwik}\ \emph {et~al.}(2003)\citenamefont {Cwik},
		\citenamefont {Lorenz}, \citenamefont {Baier}, \citenamefont {M\"uller},
		\citenamefont {Andr\'e}, \citenamefont {Bour\'ee}, \citenamefont
		{Lichtenberg}, \citenamefont {Freimuth}, \citenamefont {Schmitz},
		\citenamefont {M\"uller-Hartmann},\ and\ \citenamefont {Braden}}]{Cwik2003}%
	\BibitemOpen
	\bibfield  {author} {\bibinfo {author} {\bibfnamefont {M.}~\bibnamefont
			{Cwik}}, \bibinfo {author} {\bibfnamefont {T.}~\bibnamefont {Lorenz}},
		\bibinfo {author} {\bibfnamefont {J.}~\bibnamefont {Baier}}, \bibinfo
		{author} {\bibfnamefont {R.}~\bibnamefont {M\"uller}}, \bibinfo {author}
		{\bibfnamefont {G.}~\bibnamefont {Andr\'e}}, \bibinfo {author} {\bibfnamefont
			{F.}~\bibnamefont {Bour\'ee}}, \bibinfo {author} {\bibfnamefont
			{F.}~\bibnamefont {Lichtenberg}}, \bibinfo {author} {\bibfnamefont
			{A.}~\bibnamefont {Freimuth}}, \bibinfo {author} {\bibfnamefont
			{R.}~\bibnamefont {Schmitz}}, \bibinfo {author} {\bibfnamefont
			{E.}~\bibnamefont {M\"uller-Hartmann}},\ and\ \bibinfo {author}
		{\bibfnamefont {M.}~\bibnamefont {Braden}},\ }\bibfield  {title} {\bibinfo
		{title} {{Crystal and magnetic structure of LaTiO$_3$: Evidence for
				nondegenerate ${t}_{2g}$ orbitals}},\ }\href
	{https://doi.org/10.1103/PhysRevB.68.060401} {\bibfield  {journal} {\bibinfo
			{journal} {Phys. Rev. B}\ }\textbf {\bibinfo {volume} {68}},\ \bibinfo
		{pages} {060401} (\bibinfo {year} {2003})}\BibitemShut {NoStop}%
	\bibitem [{\citenamefont {Boysen}\ \emph {et~al.}(1976)\citenamefont {Boysen},
		\citenamefont {Ihringer}, \citenamefont {Prandl},\ and\ \citenamefont
		{Yelon}}]{Boysen76}%
	\BibitemOpen
	\bibfield  {author} {\bibinfo {author} {\bibfnamefont {H.}~\bibnamefont
			{Boysen}}, \bibinfo {author} {\bibfnamefont {J.}~\bibnamefont {Ihringer}},
		\bibinfo {author} {\bibfnamefont {W.}~\bibnamefont {Prandl}},\ and\ \bibinfo
		{author} {\bibfnamefont {W.}~\bibnamefont {Yelon}},\ }\bibfield  {title}
	{\bibinfo {title} {{X-ray and neutron investigations of the phase
				instabilities in K$_2$SnCl$_6$}},\ }\href
	{https://doi.org/https://doi.org/10.1016/0038-1098(76)90496-8} {\bibfield
		{journal} {\bibinfo  {journal} {Solid State Communications}\ }\textbf
		{\bibinfo {volume} {20}},\ \bibinfo {pages} {1019} (\bibinfo {year}
		{1976})}\BibitemShut {NoStop}%
	\bibitem [{\citenamefont {Boysen}\ and\ \citenamefont
		{Hewat}(1978)}]{Boysen78}%
	\BibitemOpen
	\bibfield  {author} {\bibinfo {author} {\bibfnamefont {H.}~\bibnamefont
			{Boysen}}\ and\ \bibinfo {author} {\bibfnamefont {A.~W.}\ \bibnamefont
			{Hewat}},\ }\bibfield  {title} {\bibinfo {title} {{A neutron powder
				investigation of the structural changes in K${\sb 2}$SnCl${\sb 6}$}},\ }\href
	{https://doi.org/10.1107/S0567740878005816} {\bibfield  {journal} {\bibinfo
			{journal} {Acta Crystallographica Section B}\ }\textbf {\bibinfo {volume}
			{34}},\ \bibinfo {pages} {1412} (\bibinfo {year} {1978})}\BibitemShut
	{NoStop}%
	\bibitem [{sup()}]{supplmat}%
	\BibitemOpen
	\href@noop {} {}\bibinfo {note} { See Supplemental Material at **** for further information about the refinements of
		the structural models with X-ray and neutron powder diffraction data.}\BibitemShut
	{NoStop}%
	\bibitem [{\citenamefont {Petříček}\ \emph {et~al.}(2014)\citenamefont
		{Petříček}, \citenamefont {Dušek},\ and\ \citenamefont
		{Palatinus}}]{Jana}%
	\BibitemOpen
	\bibfield  {author} {\bibinfo {author} {\bibfnamefont {V.}~\bibnamefont
			{Petříček}}, \bibinfo {author} {\bibfnamefont {M.}~\bibnamefont
			{Dušek}},\ and\ \bibinfo {author} {\bibfnamefont {L.}~\bibnamefont
			{Palatinus}},\ }\bibfield  {title} {\bibinfo {title} {{Crystallographic
				Computing System JANA2006: General features}},\ }\href
	{https://doi.org/doi:10.1515/zkri-2014-1737} {\bibfield  {journal} {\bibinfo
			{journal} {Zeitschrift für Kristallographie - Crystalline Materials}\
		}\textbf {\bibinfo {volume} {229}},\ \bibinfo {pages} {345} (\bibinfo {year}
		{2014})}\BibitemShut {NoStop}%
	\bibitem [{\citenamefont {Becker}\ and\ \citenamefont
		{Coppens}(1974)}]{Becker74}%
	\BibitemOpen
	\bibfield  {author} {\bibinfo {author} {\bibfnamefont {P.~J.}\ \bibnamefont
			{Becker}}\ and\ \bibinfo {author} {\bibfnamefont {P.}~\bibnamefont
			{Coppens}},\ }\bibfield  {title} {\bibinfo {title} {{Extinction within the
				limit of validity of the Darwin transfer equations. II. Refinement of
				extinction in spherical crystals of SrF${\sb 2}$ and LiF}},\ }\href
	{https://doi.org/10.1107/S0567739474000349} {\bibfield  {journal} {\bibinfo
			{journal} {Acta Crystallographica Section A}\ }\textbf {\bibinfo {volume}
			{30}},\ \bibinfo {pages} {148} (\bibinfo {year} {1974})}\BibitemShut
	{NoStop}%
	\bibitem [{Note1()}]{Note1}%
	\BibitemOpen
	\bibinfo {note} {$\beta _{\protect \rm m1} \approx \beta _{\protect \rm m2}
		\approx 90^{\circ }$, $a_{\protect \rm m2} \times \protect \sqrt {2} \approx
		b_{\protect \rm m2} \times \protect \sqrt {2} \approx c_{\protect \rm m2}$,
		$a_{\protect \rm m1} \approx b_{\protect \rm m1} \approx c_{\protect \rm
			m1}$, and $ a_{\protect \rm t} \times \protect \sqrt {2} \approx c_{\protect
			\rm t}$, where the indices $m2$, $m1$, and $t$ refer to the lattice
		parameters of the $P2_1/n$, $C2/c$, and $P4/mnc$ structural
		phases}\BibitemShut {NoStop}%
	\bibitem [{\citenamefont {Trueblood}\ \emph {et~al.}(1996)\citenamefont
		{Trueblood}, \citenamefont {B{\"{u}}rgi}, \citenamefont {Burzlaff},
		\citenamefont {Dunitz}, \citenamefont {Gramaccioli}, \citenamefont {Schulz},
		\citenamefont {Shmueli},\ and\ \citenamefont {Abrahams}}]{Trueblood96}%
	\BibitemOpen
	\bibfield  {author} {\bibinfo {author} {\bibfnamefont {K.~N.}\ \bibnamefont
			{Trueblood}}, \bibinfo {author} {\bibfnamefont {H.-B.}\ \bibnamefont
			{B{\"{u}}rgi}}, \bibinfo {author} {\bibfnamefont {H.}~\bibnamefont
			{Burzlaff}}, \bibinfo {author} {\bibfnamefont {J.~D.}\ \bibnamefont
			{Dunitz}}, \bibinfo {author} {\bibfnamefont {C.~M.}\ \bibnamefont
			{Gramaccioli}}, \bibinfo {author} {\bibfnamefont {H.~H.}\ \bibnamefont
			{Schulz}}, \bibinfo {author} {\bibfnamefont {U.}~\bibnamefont {Shmueli}},\
		and\ \bibinfo {author} {\bibfnamefont {S.~C.}\ \bibnamefont {Abrahams}},\
	}\bibfield  {title} {\bibinfo {title} {{Atomic Dispacement Parameter
				Nomenclature. Report of a Subcommittee on Atomic Displacement Parameter
				Nomenclature}},\ }\href {https://doi.org/10.1107/S0108767396005697}
	{\bibfield  {journal} {\bibinfo  {journal} {Acta Crystallographica Section
				A}\ }\textbf {\bibinfo {volume} {52}},\ \bibinfo {pages} {770} (\bibinfo
		{year} {1996})}\BibitemShut {NoStop}%
	\bibitem [{\citenamefont {Rodriguez-Carvajal}(1993)}]{Carvajal93}%
	\BibitemOpen
	\bibfield  {author} {\bibinfo {author} {\bibfnamefont {J.}~\bibnamefont
			{Rodriguez-Carvajal}},\ }\bibfield  {title} {\bibinfo {title} {{Recent
				advances in magnetic structure determination by neutron powder
				diffraction}},\ }\href
	{https://doi.org/https://doi.org/10.1016/0921-4526(93)90108-I} {\bibfield
		{journal} {\bibinfo  {journal} {Physica B: Condensed Matter}\ }\textbf
		{\bibinfo {volume} {192}},\ \bibinfo {pages} {55 } (\bibinfo {year}
		{1993})}\BibitemShut {NoStop}%
	\bibitem [{\citenamefont {Thompson}\ \emph {et~al.}(1987)\citenamefont
		{Thompson}, \citenamefont {Cox},\ and\ \citenamefont
		{Hastings}}]{Thompson87}%
	\BibitemOpen
	\bibfield  {author} {\bibinfo {author} {\bibfnamefont {P.}~\bibnamefont
			{Thompson}}, \bibinfo {author} {\bibfnamefont {D.~E.}\ \bibnamefont {Cox}},\
		and\ \bibinfo {author} {\bibfnamefont {J.~B.}\ \bibnamefont {Hastings}},\
	}\bibfield  {title} {\bibinfo {title} {{Rietveld refinement of
				Debye{--}Scherrer synchrotron X-ray data from Al${\sb 2}$O${\sb 3}$}},\
	}\href {https://doi.org/10.1107/S0021889887087090} {\bibfield  {journal}
		{\bibinfo  {journal} {Journal of Applied Crystallography}\ }\textbf {\bibinfo
			{volume} {20}},\ \bibinfo {pages} {79} (\bibinfo {year} {1987})}\BibitemShut
	{NoStop}%
	\bibitem [{\citenamefont {Finger}\ \emph {et~al.}(1994)\citenamefont {Finger},
		\citenamefont {Cox},\ and\ \citenamefont {Jephcoat}}]{Finger94}%
	\BibitemOpen
	\bibfield  {author} {\bibinfo {author} {\bibfnamefont {L.}~\bibnamefont
			{Finger}}, \bibinfo {author} {\bibfnamefont {D.}~\bibnamefont {Cox}},\ and\
		\bibinfo {author} {\bibfnamefont {A.}~\bibnamefont {Jephcoat}},\ }\bibfield
	{title} {\bibinfo {title} {{A correction for powder diffraction peak
				asymmetry due to axial divergence}},\ }\href@noop {} {\bibfield  {journal}
		{\bibinfo  {journal} {Journal of applied Crystallography}\ }\textbf {\bibinfo
			{volume} {27}},\ \bibinfo {pages} {892} (\bibinfo {year} {1994})}\BibitemShut
	{NoStop}%
	\bibitem [{Note2()}]{Note2}%
	\BibitemOpen
	\bibinfo {note} {The absorption correction is $\protect \exp (-\Sigma R)$,
		where $R$ is the radius of the cylindrical sample and $\Sigma $ is the total
		absorption length given by $ \Sigma =\protect \frac {N_f}{v_0}f\protect \frac
		{\lambda }{1.8} \sum _i c_i \sigma _{a,i}$, where $N_f$ is the number of
		formula unit, $v_0$ the volume of the unit cell at 120\protect \,K, $f=0.7$
		the estimated powder filling, $\lambda =1.494$\r A~ the neutron wavelength
		(normalized to 1.8\r A), $\sigma _{a,i}$ the absorption cross section~\cite
		{cross_sections} of element $i$ contained $c_i$ times per formula
		unit.}\BibitemShut {Stop}%
	\bibitem [{\citenamefont {Sears}(1992)}]{cross_sections}%
	\BibitemOpen
	\bibfield  {author} {\bibinfo {author} {\bibfnamefont {V.~F.}\ \bibnamefont
			{Sears}},\ }\bibfield  {title} {\bibinfo {title} {Neutron scattering lengths
			and cross sections},\ }\href {https://doi.org/10.1080/10448639208218770}
	{\bibfield  {journal} {\bibinfo  {journal} {Neutron News}\ }\textbf {\bibinfo
			{volume} {3}},\ \bibinfo {pages} {26} (\bibinfo {year} {1992})}\BibitemShut
	{NoStop}%
	\bibitem [{Note3()}]{Note3}%
	\BibitemOpen
	\bibinfo {note} {Note that in FullProf, the anisotropic ADPs are computed
		with the dimensionless $\beta _{ij}$ parameters where the indices $i,j$ refer
		to the crystallographic axis. In the manuscript, $U_{ij}$ (in~\r A$^2$) are
		reported and the following transformation is used: $U_{ij}=\beta _{ij}/(2 \pi
		^2 a^{\star }_ia^{\star }_j)$, where $a^{\star }_i$ denote the reciprocal
		lattice parameters~\cite {Trueblood96}. Note that when setting the Cl ADPs
		constraints to refine only 2 parameters, only the change of basis is
		considered in the $\beta $ calculations, but not the small
		tetragonal/monoclinic splitting and monoclinic angles, the latter having only
		a minor impact of the order $10^{-4}$ on the $U_{ij}$ values, well below the
		standard deviations.}\BibitemShut {Stop}%
	\bibitem [{\citenamefont {Koteras}\ \emph {et~al.}(2025)\citenamefont
		{Koteras}, \citenamefont {Biesenkamp}, \citenamefont {Barone}, \citenamefont
		{Mazej}, \citenamefont {Tav\ifmmode~\check{c}\else \v{c}\fi{}ar},
		\citenamefont {Hansen}, \citenamefont {Lorenzana}, \citenamefont {Grochala},\
		and\ \citenamefont {Braden}}]{Koteras25}%
	\BibitemOpen
	\bibfield  {author} {\bibinfo {author} {\bibfnamefont {K.}~\bibnamefont
			{Koteras}}, \bibinfo {author} {\bibfnamefont {S.}~\bibnamefont {Biesenkamp}},
		\bibinfo {author} {\bibfnamefont {P.}~\bibnamefont {Barone}}, \bibinfo
		{author} {\bibfnamefont {Z.}~\bibnamefont {Mazej}}, \bibinfo {author}
		{\bibfnamefont {G.~c.~v.}\ \bibnamefont {Tav\ifmmode~\check{c}\else
				\v{c}\fi{}ar}}, \bibinfo {author} {\bibfnamefont {T.~C.}\ \bibnamefont
			{Hansen}}, \bibinfo {author} {\bibfnamefont {J.}~\bibnamefont {Lorenzana}},
		\bibinfo {author} {\bibfnamefont {W.}~\bibnamefont {Grochala}},\ and\
		\bibinfo {author} {\bibfnamefont {M.}~\bibnamefont {Braden}},\ }\bibfield
	{title} {\bibinfo {title} {{Rearrangement of orbitals in
				${\mathrm{KAgF}}_{3}$ due to the Kugel-Khomskii mechanism: A neutron
				diffraction and density functional theory study}},\ }\href
	{https://doi.org/10.1103/PhysRevB.111.115156} {\bibfield  {journal} {\bibinfo
			{journal} {Phys. Rev. B}\ }\textbf {\bibinfo {volume} {111}},\ \bibinfo
		{pages} {115156} (\bibinfo {year} {2025})}\BibitemShut {NoStop}%
	\bibitem [{\citenamefont {Brown}\ \emph {et~al.}(1973)\citenamefont {Brown},
		\citenamefont {Armstrong},\ and\ \citenamefont {Jeffrey}}]{Brown73}%
	\BibitemOpen
	\bibfield  {author} {\bibinfo {author} {\bibfnamefont {A.~G.}\ \bibnamefont
			{Brown}}, \bibinfo {author} {\bibfnamefont {R.~L.}\ \bibnamefont
			{Armstrong}},\ and\ \bibinfo {author} {\bibfnamefont {K.~R.}\ \bibnamefont
			{Jeffrey}},\ }\bibfield  {title} {\bibinfo {title} {{Direct Measurement of an
				Order Parameter Associated with the 110.9-K Displacive Phase Transition in
				${\mathrm{K}}_{2}$Re${\mathrm{Cl}}_{6}$}},\ }\href
	{https://doi.org/10.1103/PhysRevB.8.121} {\bibfield  {journal} {\bibinfo
			{journal} {Phys. Rev. B}\ }\textbf {\bibinfo {volume} {8}},\ \bibinfo {pages}
		{121} (\bibinfo {year} {1973})}\BibitemShut {NoStop}%
	\bibitem [{\citenamefont {Stein}\ \emph {et~al.}(2023)\citenamefont {Stein},
		\citenamefont {Koethe}, \citenamefont {Bohat\'y}, \citenamefont {Becker},
		\citenamefont {Gr\"uninger},\ and\ \citenamefont {van Loosdrecht}}]{Stein23}%
	\BibitemOpen
	\bibfield  {author} {\bibinfo {author} {\bibfnamefont {P.}~\bibnamefont
			{Stein}}, \bibinfo {author} {\bibfnamefont {T.~C.}\ \bibnamefont {Koethe}},
		\bibinfo {author} {\bibfnamefont {L.}~\bibnamefont {Bohat\'y}}, \bibinfo
		{author} {\bibfnamefont {P.}~\bibnamefont {Becker}}, \bibinfo {author}
		{\bibfnamefont {M.}~\bibnamefont {Gr\"uninger}},\ and\ \bibinfo {author}
		{\bibfnamefont {P.~H.~M.}\ \bibnamefont {van Loosdrecht}},\ }\bibfield
	{title} {\bibinfo {title} {{Local symmetry breaking and low-energy continuum
				in ${\mathrm{K}}_{2}{\mathrm{ReCl}}_{6}$}},\ }\href
	{https://doi.org/10.1103/PhysRevB.107.214301} {\bibfield  {journal} {\bibinfo
			{journal} {Phys. Rev. B}\ }\textbf {\bibinfo {volume} {107}},\ \bibinfo
		{pages} {214301} (\bibinfo {year} {2023})}\BibitemShut {NoStop}%
	\bibitem [{\citenamefont {Dunitz}\ \emph {et~al.}(1988)\citenamefont {Dunitz},
		\citenamefont {Schomaker},\ and\ \citenamefont {Trueblood}}]{Dunitz88}%
	\BibitemOpen
	\bibfield  {author} {\bibinfo {author} {\bibfnamefont {J.~D.}\ \bibnamefont
			{Dunitz}}, \bibinfo {author} {\bibfnamefont {V.}~\bibnamefont {Schomaker}},\
		and\ \bibinfo {author} {\bibfnamefont {K.~N.}\ \bibnamefont {Trueblood}},\
	}\bibfield  {title} {\bibinfo {title} {{Interpretation of atomic displacement
				parameters from diffraction studies of crystals}},\ }\href@noop {} {\bibfield
		{journal} {\bibinfo  {journal} {The Journal of Physical Chemistry}\ }\textbf
		{\bibinfo {volume} {92}},\ \bibinfo {pages} {856} (\bibinfo {year}
		{1988})}\BibitemShut {NoStop}%
	\bibitem [{\citenamefont {Braden}\ \emph {et~al.}(2001)\citenamefont {Braden},
		\citenamefont {Meven}, \citenamefont {Reichardt}, \citenamefont
		{Pintschovius}, \citenamefont {Fernandez-Diaz}, \citenamefont {Heger},
		\citenamefont {Nakamura},\ and\ \citenamefont {Fujita}}]{Braden2001}%
	\BibitemOpen
	\bibfield  {author} {\bibinfo {author} {\bibfnamefont {M.}~\bibnamefont
			{Braden}}, \bibinfo {author} {\bibfnamefont {M.}~\bibnamefont {Meven}},
		\bibinfo {author} {\bibfnamefont {W.}~\bibnamefont {Reichardt}}, \bibinfo
		{author} {\bibfnamefont {L.}~\bibnamefont {Pintschovius}}, \bibinfo {author}
		{\bibfnamefont {M.~T.}\ \bibnamefont {Fernandez-Diaz}}, \bibinfo {author}
		{\bibfnamefont {G.}~\bibnamefont {Heger}}, \bibinfo {author} {\bibfnamefont
			{F.}~\bibnamefont {Nakamura}},\ and\ \bibinfo {author} {\bibfnamefont
			{T.}~\bibnamefont {Fujita}},\ }\bibfield  {title} {\bibinfo {title}
		{{Analysis of the local structure by single-crystal neutron scattering in
				${\mathrm{La}}_{1.85}{\mathrm{Sr}}_{0.15}{\mathrm{CuO}}_{4}$}},\ }\href
	{https://doi.org/10.1103/PhysRevB.63.140510} {\bibfield  {journal} {\bibinfo
			{journal} {Phys. Rev. B}\ }\textbf {\bibinfo {volume} {63}},\ \bibinfo
		{pages} {140510} (\bibinfo {year} {2001})}\BibitemShut {NoStop}%
	\bibitem [{\citenamefont {Kuhs}(1988)}]{Kuhs88}%
	\BibitemOpen
	\bibfield  {author} {\bibinfo {author} {\bibfnamefont {W.~F.}\ \bibnamefont
			{Kuhs}},\ }\bibfield  {title} {\bibinfo {title} {The anharmonic temperature
			factor in crystallographic structure analysis},\ }\href
	{https://doi.org/10.1071/PH880369} {\bibfield  {journal} {\bibinfo  {journal}
			{Australian Journal of Physics}\ }\textbf {\bibinfo {volume} {41}},\ \bibinfo
		{pages} {369} (\bibinfo {year} {1988})}\BibitemShut {NoStop}%
	\bibitem [{\citenamefont {Nelmes}\ \emph {et~al.}(1982)\citenamefont {Nelmes},
		\citenamefont {Meyer},\ and\ \citenamefont {Tibballs}}]{Nelmes82}%
	\BibitemOpen
	\bibfield  {author} {\bibinfo {author} {\bibfnamefont {R.~J.}\ \bibnamefont
			{Nelmes}}, \bibinfo {author} {\bibfnamefont {G.~M.}\ \bibnamefont {Meyer}},\
		and\ \bibinfo {author} {\bibfnamefont {J.~E.}\ \bibnamefont {Tibballs}},\
	}\bibfield  {title} {\bibinfo {title} {The crystal structure of tetragonal
			KH$_2$PO$_4$ and KD$_2$PO$_4$ as a function of temperature},\ }\href
	{https://doi.org/10.1088/0022-3719/15/1/005} {\bibfield  {journal} {\bibinfo
			{journal} {Journal of Physics C: Solid State Physics}\ }\textbf {\bibinfo
			{volume} {15}},\ \bibinfo {pages} {59} (\bibinfo {year} {1982})}\BibitemShut
	{NoStop}%
	\bibitem [{\citenamefont {Stokes}\ \emph {et~al.}()\citenamefont {Stokes},
		\citenamefont {Hatch},\ and\ \citenamefont {Campbell}}]{isotropysuite}%
	\BibitemOpen
	\bibfield  {author} {\bibinfo {author} {\bibfnamefont {H.~T.}\ \bibnamefont
			{Stokes}}, \bibinfo {author} {\bibfnamefont {D.~M.}\ \bibnamefont {Hatch}},\
		and\ \bibinfo {author} {\bibfnamefont {B.~J.}\ \bibnamefont {Campbell}},\
	}\href {https://iso.byu.edu/iso/isotropy.php} {\bibinfo {title} {{ISOTROPY
				Software Suite}}}\BibitemShut {NoStop}%
	\bibitem [{\citenamefont {Winter}\ \emph {et~al.}(1976)\citenamefont {Winter},
		\citenamefont {R{\"o}ssler}, \citenamefont {Bolz},\ and\ \citenamefont
		{Pelzl}}]{Winter76}%
	\BibitemOpen
	\bibfield  {author} {\bibinfo {author} {\bibfnamefont {J.}~\bibnamefont
			{Winter}}, \bibinfo {author} {\bibfnamefont {K.}~\bibnamefont {R{\"o}ssler}},
		\bibinfo {author} {\bibfnamefont {J.}~\bibnamefont {Bolz}},\ and\ \bibinfo
		{author} {\bibfnamefont {J.}~\bibnamefont {Pelzl}},\ }\bibfield  {title}
	{\bibinfo {title} {{M{\"o}ssbauer Effect and Raman Studies of the Structural
				Phase Transitions in K$_2$[SnCl$_6$]}},\ }\href@noop {} {\bibfield  {journal}
		{\bibinfo  {journal} {physica status solidi (b)}\ }\textbf {\bibinfo {volume}
			{74}},\ \bibinfo {pages} {193} (\bibinfo {year} {1976})}\BibitemShut
	{NoStop}%
	\bibitem [{\citenamefont {Pelzl}\ \emph {et~al.}(1977)\citenamefont {Pelzl},
		\citenamefont {Engels},\ and\ \citenamefont {Florian}}]{Pelzl77}%
	\BibitemOpen
	\bibfield  {author} {\bibinfo {author} {\bibfnamefont {J.}~\bibnamefont
			{Pelzl}}, \bibinfo {author} {\bibfnamefont {P.}~\bibnamefont {Engels}},\ and\
		\bibinfo {author} {\bibfnamefont {R.}~\bibnamefont {Florian}},\ }\bibfield
	{title} {\bibinfo {title} {{Raman spectroscopic study of the structural phase
				transitions in K$_2$SnCl$_6$}},\ }\href@noop {} {\bibfield  {journal}
		{\bibinfo  {journal} {physica status solidi (b)}\ }\textbf {\bibinfo {volume}
			{82}},\ \bibinfo {pages} {145} (\bibinfo {year} {1977})}\BibitemShut
	{NoStop}%
	\bibitem [{\citenamefont {Seo}\ \emph {et~al.}(1998)\citenamefont {Seo},
		\citenamefont {Pelzl},\ and\ \citenamefont {Dimitropoulus}}]{Seo98}%
	\BibitemOpen
	\bibfield  {author} {\bibinfo {author} {\bibfnamefont {Y.~M.}\ \bibnamefont
			{Seo}}, \bibinfo {author} {\bibfnamefont {J.}~\bibnamefont {Pelzl}},\ and\
		\bibinfo {author} {\bibfnamefont {C.}~\bibnamefont {Dimitropoulus}},\
	}\bibfield  {title} {\bibinfo {title} {{Comparison of 35Cl NQR Spectra
				between the Mixed Crystals K$_2$Sn$_{1-x}$Re$_x$Cl$_6$ and the Al$^{3+}$
				Doped Crystals K$_2$SnCl$_6$:Al$^{3+}$}},\ }\href
	{https://doi.org/doi:10.1515/zna-1998-6-746} {\bibfield  {journal} {\bibinfo
			{journal} {Zeitschrift für Naturforschung A}\ }\textbf {\bibinfo {volume}
			{53}},\ \bibinfo {pages} {552} (\bibinfo {year} {1998})}\BibitemShut
	{NoStop}%
	\bibitem [{\citenamefont {Ihringer}(1980)}]{Ihringer80}%
	\BibitemOpen
	\bibfield  {author} {\bibinfo {author} {\bibfnamefont {J.}~\bibnamefont
			{Ihringer}},\ }\bibfield  {title} {\bibinfo {title} {{An X-ray investigation
				of the high-temperature phase of K${\sb 2}$SnCl${\sb 6}$}},\ }\href
	{https://doi.org/10.1107/S0567739480000150} {\bibfield  {journal} {\bibinfo
			{journal} {Acta Crystallographica Section A}\ }\textbf {\bibinfo {volume}
			{36}},\ \bibinfo {pages} {89} (\bibinfo {year} {1980})}\BibitemShut {NoStop}%
	\bibitem [{\citenamefont {Lynn}\ \emph {et~al.}(1978)\citenamefont {Lynn},
		\citenamefont {Patterson}, \citenamefont {Shirane},\ and\ \citenamefont
		{Wheeler}}]{Lynn78}%
	\BibitemOpen
	\bibfield  {author} {\bibinfo {author} {\bibfnamefont {J.}~\bibnamefont
			{Lynn}}, \bibinfo {author} {\bibfnamefont {H.}~\bibnamefont {Patterson}},
		\bibinfo {author} {\bibfnamefont {G.}~\bibnamefont {Shirane}},\ and\ \bibinfo
		{author} {\bibfnamefont {R.}~\bibnamefont {Wheeler}},\ }\bibfield  {title}
	{\bibinfo {title} {{Soft rotary mode and structural phase transitions in
				K$_2$ReCl$_6$}},\ }\href
	{https://doi.org/https://doi.org/10.1016/0038-1098(78)90192-8} {\bibfield
		{journal} {\bibinfo  {journal} {Solid State Communications}\ }\textbf
		{\bibinfo {volume} {27}},\ \bibinfo {pages} {859} (\bibinfo {year}
		{1978})}\BibitemShut {NoStop}%
	\bibitem [{\citenamefont {Armstrong}(1980{\natexlab{b}})}]{Armstrong80}%
	\BibitemOpen
	\bibfield  {author} {\bibinfo {author} {\bibfnamefont {R.~L.}\ \bibnamefont
			{Armstrong}},\ }\bibfield  {title} {\bibinfo {title} {{Structural properties
				and lattice dynamics of 5d transition metal antifluorite crystals}},\ }\href
	{https://doi.org/https://doi.org/10.1016/0370-1573(80)90146-5} {\bibfield
		{journal} {\bibinfo  {journal} {Physics Reports}\ }\textbf {\bibinfo {volume}
			{57}},\ \bibinfo {pages} {343} (\bibinfo {year}
		{1980}{\natexlab{b}})}\BibitemShut {NoStop}%
	\bibitem [{\citenamefont {Kugler}\ \emph {et~al.}(1983)\citenamefont {Kugler},
		\citenamefont {Knorr},\ and\ \citenamefont {Prandl}}]{Kugler83}%
	\BibitemOpen
	\bibfield  {author} {\bibinfo {author} {\bibfnamefont {W.}~\bibnamefont
			{Kugler}}, \bibinfo {author} {\bibfnamefont {K.}~\bibnamefont {Knorr}},\ and\
		\bibinfo {author} {\bibfnamefont {W.}~\bibnamefont {Prandl}},\ }\bibfield
	{title} {\bibinfo {title} {{The lattice parameters of K$_2$SnCl$_6$ at low
				temperatures determined by high resolution single crystal X-ray
				diffraction}},\ }\href
	{https://doi.org/https://doi.org/10.1016/0038-1098(83)90700-7} {\bibfield
		{journal} {\bibinfo  {journal} {Solid State Communications}\ }\textbf
		{\bibinfo {volume} {47}},\ \bibinfo {pages} {163} (\bibinfo {year}
		{1983})}\BibitemShut {NoStop}%
	\bibitem [{\citenamefont {Ihringer}\ and\ \citenamefont
		{Abrahams}(1984)}]{Ihringer84}%
	\BibitemOpen
	\bibfield  {author} {\bibinfo {author} {\bibfnamefont {J.}~\bibnamefont
			{Ihringer}}\ and\ \bibinfo {author} {\bibfnamefont {S.~C.}\ \bibnamefont
			{Abrahams}},\ }\bibfield  {title} {\bibinfo {title} {{Soft modes and elastic
				strain at the tetragonal-to-monoclinic phase transition in antifluorite and
				related structure types}},\ }\href {https://doi.org/10.1103/PhysRevB.30.6540}
	{\bibfield  {journal} {\bibinfo  {journal} {Phys. Rev. B}\ }\textbf {\bibinfo
			{volume} {30}},\ \bibinfo {pages} {6540} (\bibinfo {year}
		{1984})}\BibitemShut {NoStop}%
	\bibitem [{Note4()}]{Note4}%
	\BibitemOpen
	\bibinfo {note} {Because of the temperature offset $\Delta $T in the PND
		measurements, in order to calculate the bond-distances and octahedral
		rotation/tilt angles, lattice parameters computed at 260\protect \,K (first
		refinement with a tetragonal cell) and 254\protect \,K are used in
		combination with the single crystal refinements at 262\protect \,K ($P4/mnc$)
		and 257\protect \,K ($C2/c$). At 150\protect \,K, the lattice parameters are
		averaged from the results obtained at 200\protect \,K and 100\protect
		\,K.}\BibitemShut {Stop}%
	\bibitem [{\citenamefont {Sperka}\ and\ \citenamefont
		{Maunter}(1988)}]{Sperka1988}%
	\BibitemOpen
	\bibfield  {author} {\bibinfo {author} {\bibfnamefont {G.}~\bibnamefont
			{Sperka}}\ and\ \bibinfo {author} {\bibfnamefont {F.~A.}\ \bibnamefont
			{Maunter}},\ }\bibfield  {title} {\bibinfo {title} {{Crystal growth and
				structure of Cs$_2$ReCl$_6$}},\ }\href
	{https://doi.org/10.1002/crat.2170230723} {\bibfield  {journal} {\bibinfo
			{journal} {Crystal Research and Technology}\ }\textbf {\bibinfo {volume}
			{23}},\ \bibinfo {pages} {K109} (\bibinfo {year} {1988})}\BibitemShut
	{NoStop}%
	\bibitem [{\citenamefont {Figgis}\ \emph {et~al.}(1961)\citenamefont {Figgis},
		\citenamefont {Lewis},\ and\ \citenamefont {Mabbs}}]{Figgis1961}%
	\BibitemOpen
	\bibfield  {author} {\bibinfo {author} {\bibfnamefont {B.~N.}\ \bibnamefont
			{Figgis}}, \bibinfo {author} {\bibfnamefont {J.}~\bibnamefont {Lewis}},\ and\
		\bibinfo {author} {\bibfnamefont {F.~E.}\ \bibnamefont {Mabbs}},\ }\bibfield
	{title} {\bibinfo {title} {{The magnetic properties of some
				$d^3$-complexes}},\ }\href {https://doi.org/10.1039/JR9610003138} {\bibfield
		{journal} {\bibinfo  {journal} {J. Chem. Soc.}\ }\textbf {\bibinfo {volume}
			{1961}},\ \bibinfo {pages} {3138} (\bibinfo {year} {1961})}\BibitemShut
	{NoStop}%
	\bibitem [{\citenamefont {Bettinelli}\ and\ \citenamefont
		{Flint}(1988)}]{Bettinelli1988}%
	\BibitemOpen
	\bibfield  {author} {\bibinfo {author} {\bibfnamefont {M.}~\bibnamefont
			{Bettinelli}}\ and\ \bibinfo {author} {\bibfnamefont {C.~D.}\ \bibnamefont
			{Flint}},\ }\bibfield  {title} {\bibinfo {title} {{Magnon sidebands and
				cooperative absorptions in K$_2$ReCl$_6$ and Cs$_2$ReCl$_6$}},\ }\href
	{https://doi.org/10.1088/0022-3719/21/32/005} {\bibfield  {journal} {\bibinfo
			{journal} {Journal of Physics C: Solid State Physics}\ }\textbf {\bibinfo
			{volume} {21}},\ \bibinfo {pages} {5499} (\bibinfo {year}
		{1988})}\BibitemShut {NoStop}%
	\bibitem [{\citenamefont {Bettinelli}\ \emph {et~al.}(1991)\citenamefont
		{Bettinelli}, \citenamefont {Flint}, \citenamefont {Ingletto}, \citenamefont
		{Veterinaria},\ and\ \citenamefont {Taglio}}]{Bettinelli1991}%
	\BibitemOpen
	\bibfield  {author} {\bibinfo {author} {\bibfnamefont {M.}~\bibnamefont
			{Bettinelli}}, \bibinfo {author} {\bibfnamefont {C.~D.}\ \bibnamefont
			{Flint}}, \bibinfo {author} {\bibfnamefont {G.}~\bibnamefont {Ingletto}},
		\bibinfo {author} {\bibfnamefont {M.}~\bibnamefont {Veterinaria}},\ and\
		\bibinfo {author} {\bibfnamefont {V.}~\bibnamefont {Taglio}},\ }\bibfield
	{title} {\bibinfo {title} {{Luminescence Properties of A$_2$ReCl$_6$
				Crystals}},\ }\href@noop {} {\bibfield  {journal} {\bibinfo  {journal} {J.
				Mater. Chem.}\ }\textbf {\bibinfo {volume} {1}},\ \bibinfo {pages} {437}
		(\bibinfo {year} {1991})}\BibitemShut {NoStop}%
	\bibitem [{\citenamefont {Braden}\ \emph {et~al.}(1994)\citenamefont {Braden},
		\citenamefont {Schweiss}, \citenamefont {Heger}, \citenamefont {Reichardt},
		\citenamefont {Fisk}, \citenamefont {Gamayunov}, \citenamefont {Tanaka},\
		and\ \citenamefont {Kojima}}]{braden1994}%
	\BibitemOpen
	\bibfield  {author} {\bibinfo {author} {\bibfnamefont {M.}~\bibnamefont
			{Braden}}, \bibinfo {author} {\bibfnamefont {P.}~\bibnamefont {Schweiss}},
		\bibinfo {author} {\bibfnamefont {G.}~\bibnamefont {Heger}}, \bibinfo
		{author} {\bibfnamefont {W.}~\bibnamefont {Reichardt}}, \bibinfo {author}
		{\bibfnamefont {Z.}~\bibnamefont {Fisk}}, \bibinfo {author} {\bibfnamefont
			{K.}~\bibnamefont {Gamayunov}}, \bibinfo {author} {\bibfnamefont
			{I.}~\bibnamefont {Tanaka}},\ and\ \bibinfo {author} {\bibfnamefont
			{H.}~\bibnamefont {Kojima}},\ }\bibfield  {title} {\bibinfo {title}
		{{Relation between structure and doping in La$_{2-x}$Sr$_x$CuO$_{4+\delta}$ a
				neutron diffraction study on single crystals}},\ }\href
	{https://doi.org/https://doi.org/10.1016/0921-4534(94)91284-X} {\bibfield
		{journal} {\bibinfo  {journal} {Physica C: Superconductivity}\ }\textbf
		{\bibinfo {volume} {223}},\ \bibinfo {pages} {396} (\bibinfo {year}
		{1994})}\BibitemShut {NoStop}%
	\bibitem [{\citenamefont {Bertin}\ \emph {et~al.}(2025)\citenamefont {Bertin},
	\citenamefont {Kiefer}, \citenamefont {Pomjakushin}, \citenamefont {Fabelo},
	\citenamefont {Becker}, \citenamefont {Bohat\'y},\ and \citenamefont {Braden}}]{zenodo-KRC}%
\BibitemOpen
\bibfield  {author} {\bibinfo {author} {\bibfnamefont {A.}~\bibnamefont
		{Bertin}}, \bibinfo {author} {\bibfnamefont {L.}~\bibnamefont {Kiefer}},
	\bibinfo {author} {\bibfnamefont {V.}~\bibnamefont {Pomjakushin}}, \bibinfo
	{author} {\bibfnamefont {O.}~\bibnamefont {Fabelo}}, \bibinfo {author}
	{\bibfnamefont {P.}~\bibnamefont {Becker}}, \bibinfo {author} {\bibfnamefont
		{L.}~\bibnamefont {Bohat\'y}}, \ and\ \bibinfo {author} {\bibfnamefont
		{M.}~\bibnamefont {Braden}},\ }\bibfield  {title}
{\bibinfo {title} {{Data and analysis codes for "Structural studies on
			$A_2$ReCl$_6$ ($A$=K, Rb, Cs): absence of Jahn-Teller distortion"}},\ }\href
{zenodo} {zenodo} (\bibinfo {year}
{2026})
\BibitemShut {NoStop}%
\end{thebibliography}
\end{document}